\documentclass[twocolumn,amssymb, amsmath, superscriptaddress,aps,showpacs,floatfix,secnumarabic,prb]{revtex4}

\usepackage{amsmath}
\usepackage{amssymb}
\usepackage{graphicx}
\usepackage{psfrag}
\usepackage{bm}
\usepackage{color}

\newcommand{\ty}[1]{\mbox{\tiny #1}}
\newcommand{\ef}{\varepsilon_{\ty{\text{F}}}}
\newcommand{\kf}{k_{\ty{\text{F}}}}
\newcommand{\vf}{v_{\ty{\text{F}}}}

\newcommand{\kb}{k_{\ty{\text{B}}}} 

\newcommand{\vs}{\vspace{0.4cm}}
\newcommand{\xie}{\xi_{\text{\tiny EC}}}

\newcommand{\xis}{\xi_{\text{\ty S}}}
\newcommand{\tgm}{\widetilde{\Gamma}}

\newcommand{\eg}{\varepsilon}
\newcommand{\z}{\mathbb{Z}}

\begin{document}
\title{Josephson current in a four terminal superconductor - exciton condensate - superconductor system}

\author{Sebastiano Peotta}
\affiliation{NEST, Scuola Normale Superiore and  Istituto Nanoscienze-CNR, I-56126 Pisa, Italy}
\author{Marco Gibertini}
\affiliation{NEST, Scuola Normale Superiore and  Istituto Nanoscienze-CNR, I-56126 Pisa, Italy}
\author{Fabrizio Dolcini}
\affiliation{Dipartimento di Fisica del Politecnico di Torino, I-10129 Torino, Italy}
\author{Fabio Taddei}
\affiliation{NEST, Istituto Nanoscienze-CNR and Scuola Normale Superiore, I-56126 Pisa, Italy}
\author{Marco Polini}
\affiliation{NEST, Istituto Nanoscienze-CNR and Scuola Normale Superiore, I-56126 Pisa, Italy}
\author{L.B. Ioffe}
\affiliation{Center for Materials Theory, Department of Physics and Astronomy, Rutgers University, 
		136 Frelinghuysen Rd, Piscataway NJ 08854 USA}
\author{Rosario Fazio}
\affiliation{NEST, Scuola Normale Superiore and  Istituto Nanoscienze-CNR, I-56126 Pisa, Italy}
\author{A.H. MacDonald}
\affiliation{Department of Physics, University of Texas at Austin, Austin, Texas 78712, USA}

\begin{abstract}
We  investigate the transport properties  of a bilayer exciton condensate that is contacted by four superconducting leads.
We focus on the equilibrium regime and investigate how the Josephson currents induced in the
bilayer by phase biases applied to the superconducting electrodes are affected by the presence of an exciton condensate in the bulk of the system. As long as the distance between the superconducting electrodes
is much larger than the exciton coherence length, the Josephson current depends 
only on the difference between the phase biases in the two layers. This result holds
true in both short- and long-junction limits. We relate it to 
a new correlated four-particle Andreev process which 
occurs at the superconductor - exciton condensate interface.  The system we investigate 
provides an implementation of  
the supercurrent mirror proposed by Kitaev as a viable way to realize topologically protected qubits.
\end{abstract}

\maketitle

\section{Introduction}
\label{sec1:level1}\noindent
 
Exciton condensates (ECs) are ordered states of matter in which macroscopic phase coherence is established through 
the condensation of electron-hole pairs. Since its early prediction in the sixties~\cite{Blatt_pr_1962,keldysh_jept_1968}
there has been considerable experimental and theoretical activity in this field.  Signatures of exciton condensation 
have been reported in quantum Hall bilayers~\cite{spielman_prl_2000}, and in 
optically-excited exciton~\cite{butov_jphys_2007} and exciton-polariton~\cite{exciton_polariton} cold gases. 

Early on it was understood~\cite{lozovik_jept_1975}  that spatially separating electrons and holes into two distinct and 
decoupled (semiconducting) layers could be extremely useful to suppress electron-hole recombination,
and in this way to enhance the possibility of realizing an exciton condensate. 
In the quantum Hall regime, exciton-condensation in bilayers can be realized in equilibrium
by condensing electrons and holes in conduction band Landau levels that are localized in separate layers~\cite{eisenstein_macdonald_nature_2004}.

When single-particle inter-layer tunneling processes can be neglected, exciton-condensation
in bilayers is equivalent to spontaneous inter-layer phase 
coherence~\cite{Kuramoto_SSC_1978,Fertig_PRB_1989,Wen_PRL_1992,Moon_PRB_1995,senatore}.  
As a result double-layer ECs support a dissipationless  ``counterflow" transport channel
 in which all electron-hole pairs drift together, giving rise to equilibrium 
counterpropagating currents in the two layers. When the two layers of a bilayer EC are separately contacted~\cite{Eisenstein_APL_1990} remarkable 
transport anomalies~\cite{spielman_prl_2000,kellogg_PRL_2004,tutuc_PRL_2004,Tiemann_NJP,Tiemann_prb_2008,finck_prl_2011},  associated with its neutral counterflow 
supercurrents~\cite{su_naturephys_2008},  are observed.  In the absence of a magnetic field, 
bilayer exciton condensation can be enabled by gating the Fermi level in one layer to the 
conduction band and the Fermi level of the other layer to the valence band.  Two experimental groups~\cite{croxall2008,seamons2009}  have recently reported the observation 
of an anomalous upturn in the Coulomb drag transresistivity as the temperature is lowered. This upturn is interpreted as being due to strong pairing fluctuations 
that precede exciton condensation~\cite{hu2000,mink_arXiv_2011} and thus serves as a precursor signal for the transition
that is similar to the enhancement of conductivity 
in superconductors due to superconducting fluctuations above but close to the critical temperature~\cite{larkin2002}.

\begin{figure}
  \includegraphics[scale=0.24]{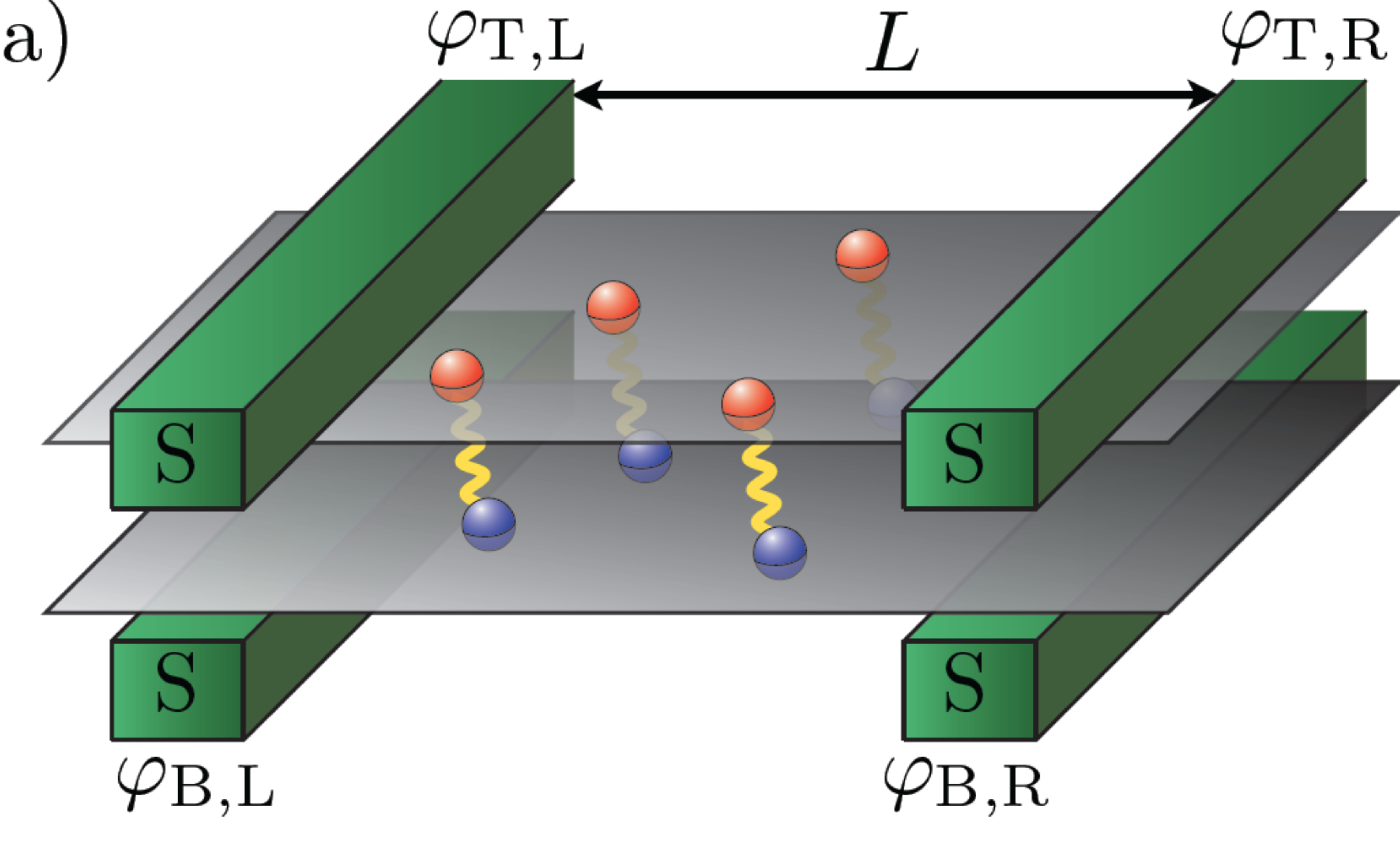}\\
  \includegraphics[scale=0.5]{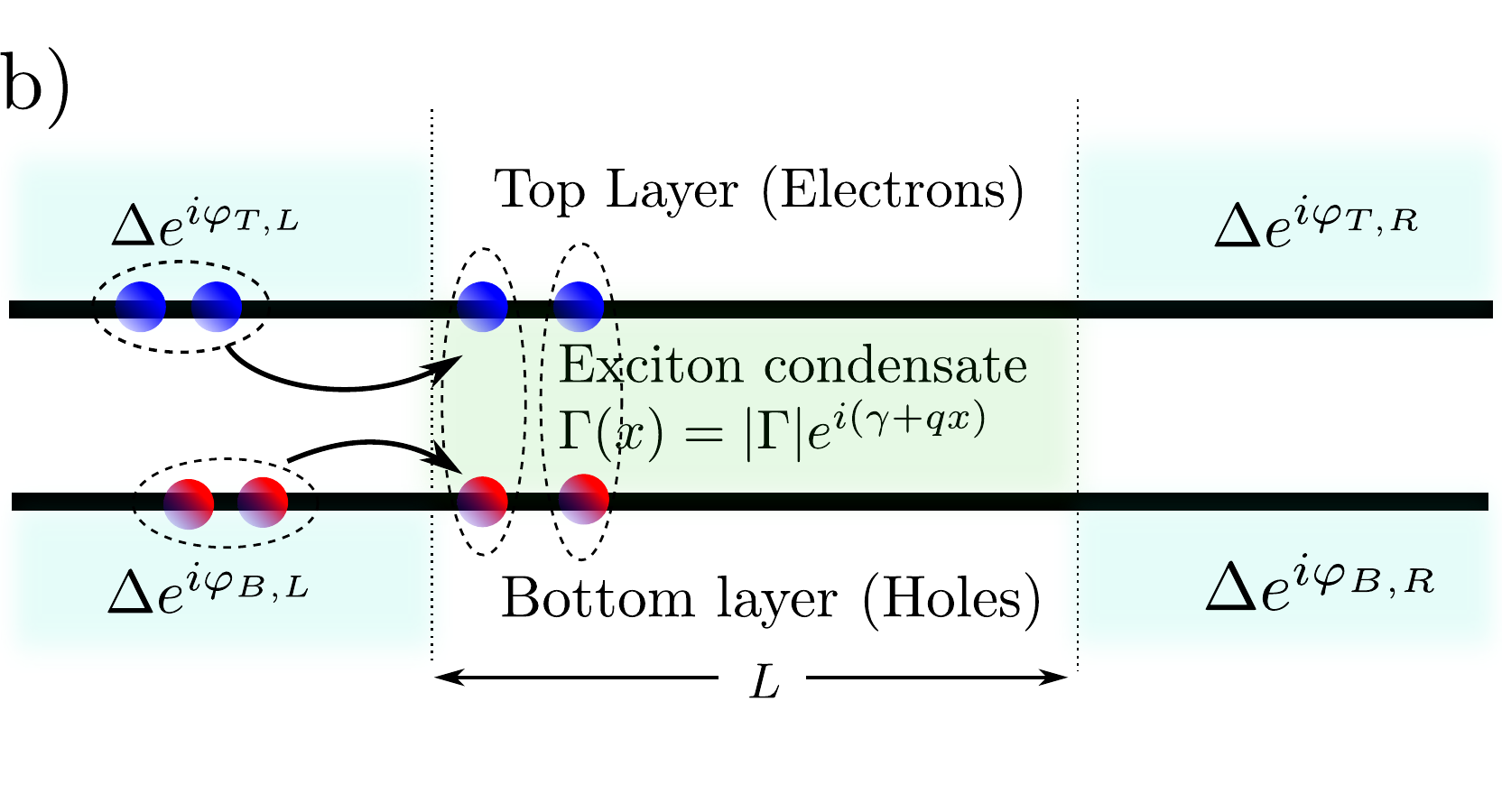}
\caption{a) Schematic of the four-terminal device studied in this work.  A double layer is coupled to four superconducting electrodes. The system is 
	phase biased and a Josephson current can flow without dissipation. The bilayer  is in an exciton condensate phase. b) Assuming translational 
	invariance along the contacts, the system can be considered as one-dimensional. In a mean-field approximation the superconducting and exciton 
	ordered phases can be described through space dependent order parameters $\Gamma(x)$ and $\Delta(x)$, respectively. The order parameter 
	$\Gamma$ is different from zero only in a region of length $L$ sandwiched  between  the electrodes. To allow for a current-currying state, the phase of the exciton condensate must 
	be space-dependent. The four superconducting electrodes have different phases $\varphi_i$.  For simplicity, we assume that the amplitude of superconducting order parameter 
	is equal to $\Delta$ in all electrodes.}
\label{setup}
\end{figure}

Despite growing experimental evidence, a definite confirmation of exciton condensation in bilayers in the 
absence of a magnetic field is still elusive.  The identification of new effects that
may highlight the physics of ECs is thus highly desirable. 
In this Article we explore hybrid phenomena in which 
counterflow superfluid currents are combined with superconducting electrodes.   
We show that when two pairs of superconducting electrodes are 
connected via a bilayer, as sketched in Fig. \ref{setup}, the Josephson current 
is dramatically influenced by exciton condensation. 
Depending on the bias configurations of the superconducting electrodes
the device can exhibit a variety of behaviors, including the {\it exciton blockade} effect, where supercurrents are suppressed, and the {\it superdrag} effect, where the system acts as a perfect current mirror. 

%%%%%%%%%%%%%%%%%%%%
%%%%%%%%%%%%%%%%%%%%
%		FIGURE
%%%%%%%%%%%%%%%%%%%%
%%%%%%%%%%%%%%%%%%%%
%%%%%%%%%%%%%%%%%%%%
%%%%%%%%%%%%%%%%%%%%
%%%%%%%%%%%%%%%%%%%%
%%%%%%%%%%%%%%%%%%%%

When a superconductor is in contact with a normal metal, 
Cooper pairs can leak across the interface.  As a result there exists a non-vanishing 
pair amplitude in the normal metal (proximity effect) which implies that 
coherence between electrons and holes in the same layer 
is induced by the coupling to the superconductor.
The proximity effect is intimately related to the microscopic mechanism of transport 
through superconductor-normal metal interfaces. 
At voltages and temperatures below the superconducting gap the dominant process is
Andreev reflection~\cite{andreev_1966}. 
An electron incoming from the normal metal  is reflected  as a hole at the interface with the superconductor, 
with consequent injection of  a 
Cooper pair into the superconductor.
Since its discovery in 1966, the study of Andreev reflection has offered many surprises. 
Of particular interest is the regime where the (incoming) 
particle and the (reflected) hole preserve their phase coherence across the normal metal~\cite{Beenakker_RMP_1997}. 
It is natural to envisage that the interplay
between phase-coherent electron propagation in metals and macroscopic phase coherence in the superconductor
will be of fundamental importance at an
EC-superconductor interface, where the neutral EC superfluid current has to convert onto a (charged) Cooper pair current. 
We will show that this 
current conversion occurs through a new kind of correlated Andreev process.  
The absorption of a Cooper pair by the exciton condensate in the 
upper layer is always accompanied by the emission of a Cooper pair in the bottom layer. 

A brief account of our results was already published in Ref.~\onlinecite{dolcini_2009}. 
Here we give details of the derivation of the published results and 
discuss new regimes not analyzed previously.  The paper is organized as follows. 
In the next Section we consider the case of an ideal interface
between an exciton condensate and a pair of superconducting electrodes.  
We define the model, introduce all the 
relevant energy scales, and discuss the correlated bilayer Andreev process 
which enables supercurrent conversion at the EC-superconductor interface.
In Sec.~\ref{longjunctionsec} we discuss the long junction limit, providing more details about the results
 presented in Ref.~\onlinecite{dolcini_2009}. 
Then in Sec.~\ref{shortjunctionsec} we introduce an alternative theoretical approach which enables us to extend our investigation of the Josephson current 
to the short junction regime. This method is based on the relationship between the scattering matrix
and the density of states.  To complete our analysis we study, in Section~\ref{sec:tunneling}, the case of poorly transmitting interfaces. In this case the Josephson current can 
be computed in a perturbation expansion in the tunneling amplitudes.  
%It is interesting to note that in this limit, one recovers a form for the phase dependent 
%term in the free energy which bears strong similarities with the one obtained in the short junction case in Sec.~\ref{shortjunctionsec}.

There is a very interesting connection between the system considered in 
this work and one put forward by Kitaev~\cite{kitaev} to realize topological protected qubits.
In Sec.~\ref{protected} we will explore the similarities and 
discuss to which extent the system considered in this work can implement topological quantum computation. 
In the concluding section we summarize the main results of 
our work and comment on the feasibility of testing our findings experimentally in semiconductor or graphene double layers.  Technical details of most derivations have been relegated to Appendices.

\section{Josephson current for ideal superconductor-exciton condensate interfaces}
\label{sec:model}
In the Josephson effect~\cite{barone_paterno} a supercurrent can flow through a weak link in the absence of an 
electrical bias whenever a phase bias is present.  In the usual two-terminal setup, the weak link can be a tunnel barrier 
(such as a normal metal or a semiconductor sandwiched between two superconductors) or any sort of constriction.
In this work we consider~\cite{dolcini_2009} two coupled weak links between {\it four} superconducting electrodes,
linked by a bilayer of length $L$, as depicted in Fig.~\ref{setup}.
The electron and hole densities in the two layers can be tuned separately through external gates. 
We assume that the top ($T$) layer is 
electron-doped, {\em i.e.} negatively charged, and the bottom ($B$) layer is hole-doped, {\em i.e.} positively charged. 
The two layers are coupled only through the Coulomb interaction.  (Direct single-electron tunneling between the layers is
assumed to have been suppressed by a dielectric barrier.)

\subsection{The Hamiltonian}

Assuming translational invariance along the junctions, the problem to be considered is one-dimensional (longitudinal modes with 
different transverse momentum  are not coupled). 
%%%%%%%%%%%%%%%%%%%%
%%%%%%%%%%%%%%%%%%%%
%	FIGURE
%%%%%%%%%%%%%%%%%%%%
%%%%%%%%%%%%%%%%%%%%
 \begin{figure}
 \includegraphics[scale=0.30]{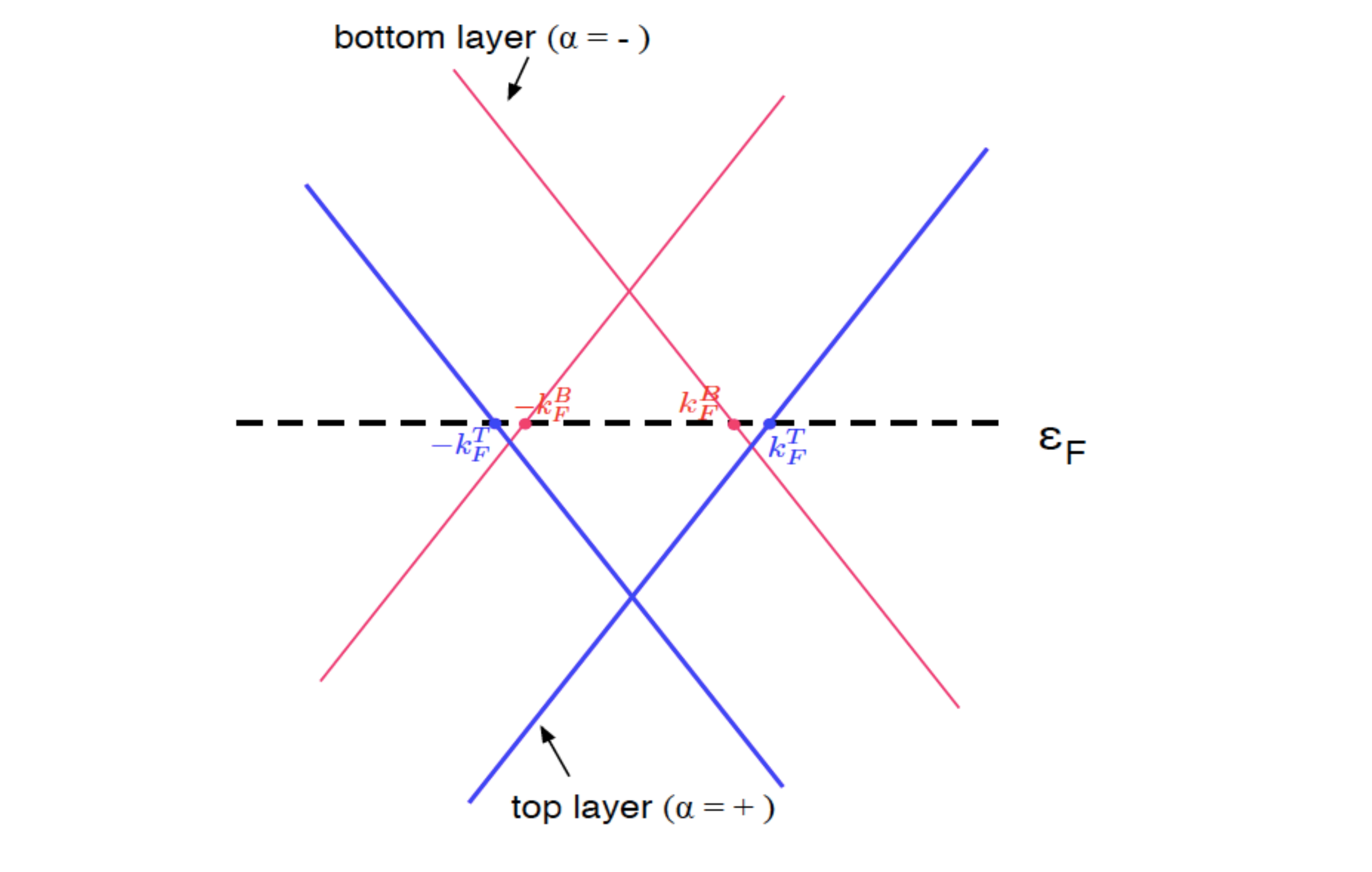}
 \caption{One-dimensional linearized bands for the two layers. For a Fermi level location
 in the figure, top (bottom) layer states near the right Fermi point are right-movers (left-movers), 
 whereas states near the the left Fermi point are left-movers (right-movers)}
 \label{bands}
 \end{figure}
%%%%%%%%%%%%%%%%%%%%
%%%%%%%%%%%%%%%%%%%%
%%%%%%%%%%%%%%%%%%%%
%%%%%%%%%%%%%%%%%%%%
At low energies the dispersion relation can be linearized around the Fermi energy $\varepsilon_{\text{\tiny F}}$. 
In this large Fermi-energy limit terms of ${\cal O}(\Delta/\varepsilon_{\text{\tiny F}})$ 
and ${\cal O}(\Gamma/\varepsilon_{\text{\tiny F}})$ are automatically set to zero, where $\Delta$ and $\Gamma$ 
are the superconducting and EC order parameters, respectively (see below). In the literature the approximation in which these terms are neglected is usually termed 
``Andreev approximation"~\cite{bardeen_johnson_1972,ishii_1970,kulik_1970,andreev_1966}.

After linearization the spectrum is split into two distinct branches denoted by the two-valued quantum number $p=\pm$.
On the $T$ ($B$) layer excitations with $p = +$ correspond to right (left) movers and 
those with $p = -$ to left (right) movers (see Fig.~\ref{bands}).  The electron field operator of layer $\alpha = T, B$ and spin component $\sigma$ 
can be written as
\begin{equation}\label{fermifield}
	\Psi_{\alpha \sigma}(x) = 
	e^{i k_{\text{\tiny F}}^{(\alpha)} x} \Psi_{\alpha \sigma +}(x)+ e^{-i k_{\text{\tiny F}}^{(\alpha)} x} \Psi_{\alpha \sigma -}(x)
\end{equation}
where $\Psi_{\alpha \sigma \pm}(x)$ are fields related to the Fermi point $\pm k_{\text{\tiny F}}^{(\alpha)}$ 
that are assumed to be slowly varying over the length scales 
$\lambda_F^\alpha =2 \pi/k_\text{\tiny F}^{(\alpha)}$.

At the mean-field level the exciton and superconducting condensates are described 
by the two order parameters $\Gamma (x) \propto  \langle 
\Psi^{\dagger}_{B \sigma}(x)  \Psi^{}_{T \sigma}(x) \rangle$ and  $\Delta(x) \propto  \langle \Psi_{\alpha \downarrow}(x) \Psi_{\alpha \uparrow}(x)  \rangle$, respectively.
The EC pairing potential pairs holes in the bottom layer and electrons in the top layer, while  the superconducting order parameter 
pairs right-moving (left-moving) electrons with spin up and left-moving (right-moving) electrons with spin down separately in each layer. 
Here we assume that the bilayer coherence is between up spins and up spins and 
equivalently between down spins and down spins.  In principle there is a set of equivalent energy 
bilayer states related to these by spin-rotation in one layer only.
The mean-field Hamiltonian $\mathcal{\hat{H}}$, quadratic in the fermion operators,  which 
captures interactions with these condensates has a  (generalized) Bogoliubov-de~Gennes form~\cite{degennes_1966}.
In Nambu notation it reads
$$
	\mathcal{\hat{H}}=\int_{-\infty}^{+\infty} dx \sum_{p=\pm}\Psi_p^{\dagger}(x)\mathcal{H}_p\Psi_p(x) \;\; ,
$$
where
\begin{equation}
\label{eq:ham}
 \mathcal{H}_p = \begin{pmatrix}
                -ip\hbar\vf\partial_x & \Gamma(x) & \Delta_{T}(x) & 0 \vspace{0.4cm} \\
                \Gamma^*(x) & ip\hbar\vf\partial_x & 0 & \Delta_{B}(x) \vspace{0.4cm}\\
                \Delta_{T}^*(x) & 0 & ip\hbar\vf\partial_x & -\Gamma^*(x) \vspace{0.4cm}\\
                0 & \Delta_{B}^*(x) & -\Gamma(x) & -ip\hbar\vf\partial_x
               \end{pmatrix}~,
\end{equation}
($\vf$ being the Fermi velocity) and
\begin{equation}
	\Psi_p(x) = \big(\Psi_{T\uparrow p}(x),\Psi_{B\uparrow p}(x),
	\Psi^{\dagger}_{T\downarrow \bar{p}}(x),\Psi^{\dagger}_{B\downarrow\bar{p}}(x) \big)^\mathrm{T} \, .
\end{equation}
In principle the two order parameters have to be determined self-consistently. 
In the regimes considered in this paper, however, self-consistency would 
introduce only negligible quantitative changes of the results. A more detailed discussion on the situations in which self-consistency is unimportant and the order parameters 
can be taken to have a step-like form is discussed in Ref.~\onlinecite{beenakker_transport}.  The EC order parameter $\Gamma(x)$ is assumed to be 
uniform in amplitude in the EC region ($-L/2\leq x \leq L/2$) and zero  otherwise.  
In order to account for the neutral  counterflow current, the phase of the condensate is allowed to be
space-dependent. Current conservation in one dimension for an order parameter
with a constant amplitude implies the following functional dependence on $x$:
\begin{equation}\label{Gamma_x_def}
 \Gamma(x) = |\Gamma|e^{i(\gamma+qx)}~,  
\end{equation}
where the wave vector $q$ and the phase $\gamma$ are to be fixed by free-energy minimization. 
The superconducting order parameter, on the other hand, is assumed to be 
different from zero only in the electrodes:
\begin{gather}
 \Delta_{T}(x) = \begin{cases}
                          \Delta e^{i\varphi_{T,L}}, & x< -L/2\\
			  0, & -L/2 \leq x\leq L/2\\
			  \Delta e^{i\varphi_{T,R}}, & x>L/2
                         \end{cases}
\end{gather}
and
\begin{gather}
\Delta_{B}(x) = \begin{cases}
                 \Delta e^{i\varphi_{B,L}}, & x <-L/2\\
		0, & -L/2\leq x\leq L/2\\
		 \Delta e^{i\varphi_{B,R}}, & x> L/2
                \end{cases}~.
\end{gather}
For the sake of convenience, we have assumed that only the phases vary from one superconductor to the other.
Because charge is conserved separately in the two layers, the number of independent 
phases is reduced to {\it two}, which we take to be the phase differences between left and right electrodes 
in the top and bottom layers. 

Because the potentials are constant and are never 
simultaneously non-zero, it is easy to solve the Bogoliubov-de Gennes equations
separately in the excitonic and superconducting regions. (See Appendix~\ref{app:sol}.)
In addition to the distance between the electrodes, $L$, the 
other length scales which must be considered in coupling 
different regions are the exciton and superconducting coherence lengths, $\xie = \hbar \vf/|\Gamma|$ and $\xis = \hbar \vf/(\pi \Delta)$, 
and, at finite temperature, the thermal length $L_{\rm th}=\beta \hbar\vf $ [with $\beta= (\kb T)^{-1}$]. 

The Josephson current in this four-terminal system has two contributions.
The first one is related to direct quasi-particle tunneling through the barrier caused by the gap in the 
bilayer quasi-particle spectrum. 
This process is present in ordinary Superconductor-Insulator-Superconductor 
two-terminal setups. The second contribution, which is the main subject of interest here, is linked to the possibility of direct dissipationless conversion of supercurrents into EC counterflow currents
and can be realized only in a four-terminal setup like the one shown 
in Fig.~\ref{setup}.  Importantly, in the limit $L \gg \xie $ the first contribution is 
exponentially suppressed and only the second one survives. 
In this case, equilibrium properties will ultimately depend only 
on {\it one} phase variable. We will restrict ourselves to this regime, and address the following two distinct limits:
\begin{eqnarray}
\mbox{Long-junction limit} \hspace{1cm}  \;\;\;\;\; L \gg \xis~,
\label{longj} \\
\mbox{Short-junction limit} \hspace{1cm} \;\;\;\;\; L \ll \xis~.
\label{shortj}
\end{eqnarray}
We will further extend our results to the ``tunneling limit" ({\it i.e.} the regime in which the contacts have a low transparency) 
in which the Josephson effect was originally studied. Our results in this case are qualitatively similar to the ones in the ballistic regime.

\subsection{Four-particle Andreev reflection }

The conversion of supercurrent into counterflow (neutral) current 
hidden in the solution of the Bogoliubov-de Gennes equations, takes place through a
coherent Andreev-like process which involves \emph{two} Cooper pairs in the two different layers.  
We find it useful to give a brief illustration of this 
process before going into the details of our results for the Josephson current. 
A detailed derivation will be presented in the following Sections and in the Appendices.

\begin{figure}
\begin{center}
\includegraphics[scale=0.7]{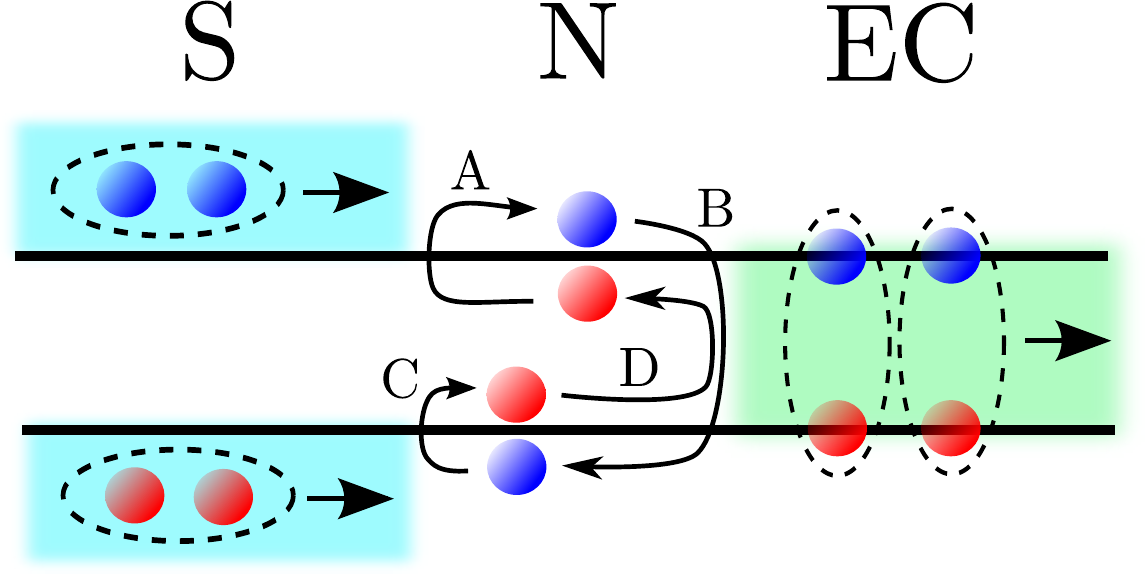} 
 \caption{\label{car}Four-particle coherent Andreev reflection. -- This illustration is a schematic for 
 the process responsible for the conversion of supercurrent from the two superconductors 
into an excitonic counterflow supercurrent. An artificial 
normal (N) region in which both superconducting and EC pairing potentials are zero has been inserted between the 
superconductors (S)  and the exciton condensate (EC). 
At the superconductor-normal interface an electron impinging from the normal region is perfectly reflected as a hole (arrow C) 
or, viceversa, a hole is reflected as  an electron (arrow A). 
Current is conserved by emission/absorption of a Cooper pair. This is the normal Andreev reflection process.
At the right side a similar but distinct process takes place. 
The exciton pairing potential has the effect of reflecting electrons and holes from one layer to the other
(arrows B and D).  Since real interlayer tunneling is absent 
the interlayer reflection is always accompanied by the emission of an exciton 
(electron-hole pair). Two excitons are emitted, one for the process B, and one for D. 
The wavefunctions for the complete path ABCD can be matched 
only at a precise energy which depends on the various phases at play 
and is the energy of the excitonic Andreev bound state (EABS) mentioned in the text.
The process illustrated converts  Cooper pairs in the two superconductors into two excitons.}
\end{center}
\end{figure}

It is sufficient to consider only one superconductor - EC interface
as shown in Fig.~\ref{car}.  It is useful to imagine a thin normal region between the superconducting and EC regions 
in which the bilayer is in its normal phase with {\it both} order parameters set to zero.
We limit our attention here to energies smaller 
than the EC and superconducting gaps.  
As already discussed, the wave function has four distinct amplitudes $\Psi = (\psi_{T\uparrow},\psi_{B\uparrow},\psi_{T\downarrow},\psi_{B\downarrow})$.
The superconducting order parameter is responsible 
for the Andreev reflection process in which an electron impinging on the superconductor
from the normal region is reflected as a hole and overall current is 
conserved because an electron Cooper pair is created (or equivalently a hole Cooper pair is destroyed). 
This process is illustrated on the left side of 
the bottom layer in Fig.~\ref{car}. 
The time reversed process is obviously allowed as well and 
is illustrated on the top left of Fig.~\ref{car}. For the wave function this implies 
the following two relations between up and down spin amplitudes in the same layer:
\begin{equation}
 \psi_{T\uparrow} = a~\psi^*_{T\downarrow}, \qquad \psi_{B\downarrow}= b~\psi^*_{B\uparrow}\,. \label{And-refl}
\end{equation}
where $a,b$ are complex number with $|a| = |b| = 1$.  Explicit expressions for $a$ and $b$
(as well as for the quantities $c$ and $d$ defined below) as functions of the energy will be calculated later in the paper.
An analogous process takes place at the normal-EC interface.
As shown on the right hand side of Fig.~\ref{car}, a particle (either electron or hole) impinging on the EC from 
the top layer is reflected into the bottom layer upon emission (or absorption) of an electron-hole pair (exciton). The same process allows a particle to go from the bottom to the top layer.
The following two additional relations between top and bottom layer amplitudes capture these processes
\begin{equation} \label{exc-refl}
\psi_{T\downarrow} = c~\psi_{B\downarrow},\qquad \psi_{B\uparrow} = d~\psi_{T\uparrow}~,
\end{equation}
where $c$ and $d$ are complex numbers with $|c| = |d| =1$. Combining the four amplitude relations we conclude that
\begin{equation}\label{eq}
 \psi_{T\uparrow} = ac^*b^*d~\psi_{T\uparrow}\Rightarrow ac^*b^*d = 1~.
\end{equation}
Solving Eq.~\eqref{eq} determines the 
energy of a state that is bound to an interface by the gaps in both superconducting and 
excitonic regions.  We refer to this state as to an \emph{Excitonic Andreev Bound State} (EABS in the following) 
It is well known that Andreev bound  states in Superconductor - Normal metal - Superconductor systems~\cite{bardeen_johnson_1972,ishii_1970} are  responsible for the flow of Josephson current. 
Although they differ in origin and in properties, EABS will 
also be crucial for supercurrent conversion from the contacts to the exciton condensate.  
Indeed, using general arguments based on current conservation,
it is natural to expect that a phase bias applied between 
the top or bottom superconducting electrodes will result in a Josephson current
equal in magnitude but {\em opposite } in sign in the two layers. 
This must be true since only counterflow currents can flow deep in the EC.
When the total energy of the system is minimized to find an equilibrium state, the 
counterflow current generated at the left junction must be absorbed at the right.
The total Josephson current then turns out to depend only on the {\em difference of the phase differences} across the 
top and the bottom layers, as we show explicitly below.
%%%%%%%%%%%%%%%%%

%%%%%%%%%%%%%%%%%%%%
%%%%%%%%%%%%%%%%%%%%
%		FIGURE
%%%%%%%%%%%%%%%%%%%%
%%%%%%%%%%%%%%%%%%%%
%%%%%%%%%%%%%%%%%%%%
%%%%%%%%%%%%%%%%%%%%
%%%%%%%%%%%%%%%%%%%%
%%%%%%%%%%%%%%%%%%%%

%
\subsection{Long-junction limit}
\label{longjunctionsec}

We start our analysis from the long-junction limit, Eq.~\eqref{longj}, where the superconducting gap is the largest energy scale in the problem. The effect of the superconducting leads can be accounted for by means of proper boundary conditions. Indeed in the deep subgap energy regime $\eg \ll \Delta$, electrons impinging at the interface 
with the superconductor are totally Andreev reflected with a reflection coefficient that is energy independent.  The boundary conditions at the two interfaces $x=0$ and $x=L$ for the left and right moving fields [introduced in Eq.~\eqref{fermifield}] were derived by Maslov {\em et al.}~\cite{maslov} and, for the present case, they read
\begin{gather}
	 \hat{\Psi}_{\alpha\uparrow+}(0) = -i\alpha e^{i\varphi_{\alpha,L}}\hat{\Psi}^{\dagger}_{\alpha\downarrow-}(0)~, \label{eq:cond1}\\
	 \hat{\Psi}_{\alpha\downarrow+}(0) = +i\alpha e^{i\varphi_{\alpha,L}}\hat{\Psi}^{\dagger}_{\alpha\uparrow-}(0)~, \label{eq:cond2}\\
	 \hat{\Psi}_{\alpha\uparrow+}(L) = +i\alpha e^{i\varphi_{\alpha,R}}\hat{\Psi}^{\dagger}_{\alpha\downarrow-}(L)~, \label{eq:cond3}\\
	 \hat{\Psi}_{\alpha\downarrow+}(L) = -i\alpha e^{i\varphi_{\alpha,R}}\hat{\Psi}^{\dagger}_{\alpha\uparrow-}(L)~. \label{eq:cond4}
\end{gather}
A convenient way to implement these boundary conditions is to perform a {\it field folding}, 
namely to express the fields $\hat{\Psi}^{\dagger}_{\alpha\bar{\sigma}-}$ in terms of 
$\hat{\Psi}_{\alpha\sigma+}$ defined on an extension of the system to coordinate $-L \leq x \leq 0$. More specifically one defines, for $x \geq 0$
\begin{gather}
 \hat{\Psi}_{\alpha\uparrow+}(-x) \equiv -i\alpha e^{\varphi_{\alpha,L}}\hat{\Psi}^{\dagger}_{\alpha\downarrow-}(x)~, \\
 \hat{\Psi}_{\alpha\downarrow+}(-x) \equiv +i\alpha e^{\varphi_{\alpha,L}}\hat{\Psi}^{\dagger}_{\alpha\uparrow-}(x)~.
\end{gather}
The first two conditions \eqref{eq:cond1} and \eqref{eq:cond2} at $x = 0$ are automatically satisfied by requiring the continuity of the new
fields. These definitions in turn imply that
\begin{equation}\label{eq:twisted}
 \hat{\Psi}_{\alpha\sigma+}(-L) = -e^{i(\varphi_{\alpha,L}-\varphi_{\alpha,R})}\hat{\Psi}_{\alpha\sigma+}(L)~.
\end{equation}
For a given layer and spin direction, the two-field model ($p=\pm$) can thus be mapped into a one-field model $\hat{\Psi}_{\alpha\sigma+}$ defined
on the interval $-L \leq x \leq L$, and satisfying the \emph{twisted} boundary condition \eqref{eq:twisted}.
Consequently the free Hamiltonian can be rewritten in the interval $-L<x<L$ in the form
\begin{equation}
 	\hat{\mathcal{H}}_{\alpha} = \int_{-L}^Ldx\, \sum_{\sigma= \uparrow,\downarrow}\left[-i\alpha\hbar\vf\hat{\Psi}_{\alpha\sigma+}^\dagger(x)\partial_x
 	\hat{\Psi}_{\alpha\sigma+}(x)\right]\,. \label{H-alpha-folded}
\end{equation}
Similarly, after the folding transformations \eqref{eq:cond1} and \eqref{eq:cond2} the excitonic coupling term  reads:
\begin{equation}
 	\hat{\mathcal{H}}_{\Gamma} 
 	= \sum_{\sigma = \uparrow,\downarrow} \int_{-L}^{L}dx\,\widetilde{\Gamma}(x)\left[\hat{\Psi}^{\dagger}_{T\sigma+}(x)
 \hat{\Psi}_{B\sigma+}(x)+\mathrm{H.c.}\right]\,,\label{H-Gamma-folded}
\end{equation}
where the folded EC order parameter reads
\begin{equation}
 	\widetilde{\Gamma}(x) = \left\{\begin{array}{ll}
	|\Gamma|e^{i\gamma+iqx} & x > 0\,, \\
	|\Gamma|e^{-i(\gamma+qx+\varphi_{T,L}-\varphi_{B,L})} & x < 0\,.
         \end{array}\right.
\end{equation}

\noindent

The Josephson current through the $\alpha=\pm=T,B$ layer can be evaluated as
\begin{equation}
\label{eq:current_op}
	\langle I_\alpha \rangle  =  \alpha \, e   \vf   \sum_{\sigma=\uparrow, \downarrow, p=\pm} p \langle    
	\, \Psi^\dagger_{\alpha \sigma  p}(x)  \Psi_{\alpha \sigma p}(x) \rangle \;\; .    
\end{equation}
Notice that right-movers are characterized by $\alpha p =+1$, so in the top layer their momentum is located near the right Fermi point 
$+k_{\text{\tiny F}}$, while in the bottom layer is located near the left Fermi point $-k_{\text{\tiny F}}$, as shown also in Fig.~\ref{bands}. Similarly left-movers are characterized by $\alpha p =-1$.
Since all the electrodes are at the same chemical potential (no voltage bias is applied), the only contribution to the current is due to the Josephson term.
The   folded Hamiltonian $\hat{\mathcal{H}}=\hat{\mathcal{H}}_T+\hat{\mathcal{H}}_B+\hat{\mathcal{H}}_\Gamma$ [see Eqs.~(\ref{H-alpha-folded})-(\ref{H-Gamma-folded})] can be straightforwardly diagonalized in the space defined by the 
boundary condition~(\ref{eq:twisted}). The supercurrent is then  evaluated using Eq.~\eqref{eq:current_op} and can be expressed  (details are provided in Appendix~\ref{folded-derivation}) as the sum of a ground state contribution and a thermal fluctuation term~\cite{dolcini_2009}:
\begin{equation}
	I_\alpha  = I_{\alpha,  GS} + I_{\alpha, TF}~.
\end{equation}
\begin{widetext}
The ground state current reads
\begin{equation}\label{GS-current}
 	I_{\alpha,GS}= -\alpha \, \frac{e  v_F}{L}  \lim_{y \rightarrow x} \left[    \sum_{k^{(0)}}  
	\left( e^{i  \alpha q (x-y)}  F(k^{(0)}+\alpha   \bar{k})  e^{i \alpha (k^{(0)}+\alpha \bar{k})(x-y)} \, \,   
	 - e^{-i  \alpha q (x-y)} \, F(k^{(0)}-\alpha \bar{k})  e^{i  \alpha (k^{(0)}-\alpha \bar{k})(x-y)}   \, \right)
	\,  \right] 
\end{equation}
while the  thermal fluctuation current is given by
\begin{equation}
\label{TF-current}
I_{\alpha,TF} = \displaystyle - 2   \, \frac{e  v_F}{L}
\sum_{k^{(0)}} \left\{ \frac{F(k^{(0)}- \bar{k})}{1+ e^{\beta 
	\left(  -\alpha \hbar v_F q +\sqrt{|\Gamma|^2+(\hbar v_F (k^{(0)}- \bar{k}))^2}\right) }} -
\frac{F(k^{(0)}+ \bar{k})}{1+ e^{ \beta \left( \alpha \hbar v_F q +\sqrt{|\Gamma|^2+
	(\hbar v_F (k^{(0)}+ \bar{k}))^2}\right)}}  \right\}  ~ .
\end{equation}
\end{widetext}
In the above equations  we have defined
\begin{equation}\label{k-vari}
	k^{(0)}=\frac{(2n + 1)\pi}{2L} \hspace{1cm}  \bar{k} =\frac{\varphi_T+\varphi_B +2 \pi J}{4L} 
\end{equation}
with $n$ and $J$ relative integers. The EC phase-winding wave vector is fixed by the condition
\begin{equation}\label{eq:qll}
	qL = \frac{\varphi_T-\varphi_B}{2}+n\pi~,
\end{equation}
where $n$ is a relative integer and 
\begin{eqnarray}
\varphi_{T} & \equiv &\varphi_{T,R}-\varphi_{T,L}~, \label{varphiT-def} \\
\varphi_{B} & \equiv &\varphi_{B,R}-\varphi_{B,L}~.  \label{varphiB-def} 
\end{eqnarray}
Finally the function $F(k)$ is defined as
\begin{equation}
	F(k)     =  \frac{1}{2} \frac{\sqrt{1+(k \xie)^2}+ k \xie }{\sqrt{1+(k \xie)^2}} ~,
\label{F-def}  
\end{equation}
{\it i.e.} it is a Heaviside-like  function, smoothed over a length~$\xie$.
We observe that $F(k)$ behaves as
\begin{displaymath}
F(k) \sim \frac{1}{4 \, (k \xie)^2 } \hspace{0.2cm} \mbox{for} \, \, \, k \rightarrow -\infty
\end{displaymath}
while for $|\Gamma| \rightarrow 0$ tends to the Heaviside step function, 
$F(\alpha k) \rightarrow \theta(\alpha k)$.

We recall that the superconducting gap has been set as the largest energy scale in the problem and thus it does not appear in the expression 
for the current given in Eqs. \eqref{GS-current} and \eqref{TF-current}. In the following we will discuss the properties of the Josephson 
current at zero and finite temperatures.

\subsubsection{Zero temperature}

In the case of a long junction the function $F(k)$ defined in (\ref{F-def}) varies smoothly with respect to the discrete $k^{(0)}$ spectrum and one can transform the sum in Eq.~(\ref{GS-current}) into an integral.   The counterflow current can be related to $q$ which depends solely on the difference of the two phase differences because of the condition given in Eq.~(\ref{eq:qll}). The resulting expression, given in Ref.~\onlinecite{dolcini_2009}, is
\begin{equation}\label{eq:to}
	I_{T/B} = \pm\frac{e\vf}{L}\left(\frac{\varphi_T-\varphi_B}{2\pi}\right)\,,\qquad \varphi_T-\varphi_B \in [\,-\pi;\,\pi]\,.
\end{equation}
Eq.~(\ref{eq:to}) has the form anticipated previously from general arguments. 
As we will show in Sections~\ref{results}  - \ref{sec:ballistic-short}, it is possible to obtain the same result by energy 
minimization. 

The Josephson current $I_{T/B}$ depends on the difference between the two phase differences (top and bottom). This peculiar phase dependence has several interesting physical implications. When the top and bottom junctions are polarized with the same 
phase bias ($\varphi_T = \varphi_B$, parallel flow) no supercurrents can flow through the EC. In this case the Josephson 
currents experience an {\em exciton blockade}. In the opposite case of counterflow phase bias ($\varphi_T = -\varphi_B$) the Josephson 
current flowing through the EC is maximal, with a  critical value equal to {\it half} of the critical current of a ballistic one-channel Superconductor - Normal metal - Superconductor (S-N-S) junction.
This is evidence of the fact that the four-terminal device allows supercurrent  (dissipationless) drag.  When current flows 
in one layer due to a phase bias in the same layer, a current equal in magnitude but opposite in direction flows in the other layer. 
This is a consequence of the perfect conversion of exciton current into supercurrent.  Eq.~\eqref{eq:to} can then be seen as a perfect drag effect for the supercurrent. 

\subsubsection{Finite temperature}

The contribution to the current due to thermal fluctuations is given by Eq.~\eqref{TF-current}. It depends both  on the {\em sum} $\varphi_T+\varphi_B$ and on $q$ 
[{\it i.e.} on the {\em difference} $\varphi_T-\varphi_B$, see Eq.~\eqref{eq:qll}]. As previously mentioned, under the condition $\xie \ll L$, one can fairly well approximate the 
sum with an integral and the resulting expression is a function of $q$ only. 
Furthermore, under the condition $ \beta |\Gamma| \gg 1 $ one can approximate the Fermi functions in Eq. \eqref{TF-current} with exponentials obtaining the following expression for the total current
\begin{equation}\label{eq:finite_temperature_current}
	I_{T/B} = \pm\frac{e\vf}{\pi}q\left[1-\sqrt{2\pi\beta|\Gamma|} 
	\frac{\sinh\left(qL_{\text{th}}/2\right)}{qL_\text{th}/2}e^{-\beta|\Gamma|}\right]\,,
\end{equation}
valid in the relevant regime $\hbar\vf/L \ll k_{\text{\tiny B}}T \ll |\Gamma|$, with $q$ fixed by Eq.~\eqref{eq:qll}. 
The first term in square brackets is 
the $T=0$ contribution, Eq.~\eqref{eq:to}, while the second one encodes the effect of thermal fluctuations 
and scales as $e^{-\beta|\Gamma|}$. This 
means that as long  as thermal fluctuations are dominated by the exciton gap, the ground-state current is essentially unaffected by finite temperatures.  Note that  this occurs even when the thermal length $L_{\text{th}}$ is smaller than
the length $L$ of the junction. This is in striking 
contrast to the case of an S-N-S junction (or with the case of two decoupled layers), where the critical current is exponentially suppressed\cite{bardeen_johnson_1972}. In the presence of the EC, Andreev reflection processes 
occurring coherently at the two interfaces transform Cooper pairs into the electron-hole 
pairs of the EC, which are protected from thermal decoherence by the excitonic gap.  
Thus in the temperature window $\hbar\vf/L \ll k_{\text{\tiny B}}T
 \ll |\Gamma|$ the EC counterflow channel is responsible for an \emph{exponential enhancement} of the critical  current. 
This effect should be readily observable as an anomalous persistence of the saw-tooth Josephson current as temperature is increased.

%%%%%%%%%%%%%%%%%%%%%%%%%%%%%%%%%%%%%%%%%%%%%%%%%%%%%%%%%%%%%%%%%%%%%%%%%%%%%%%%%%%%%%%%%%%%%%%%%%%%%%
%%%%%%%%%%%%%%%%%%%%%%%%%%%%%%%%%%%%%%%%%%%%%%%%%%%%%%%%%%%%%%%%%%%%%%%%%%%%%%%%%%%%%%%%%%%%%%%%%%%%%%
%%%%%%%%%%%%%%%%%%%%%%%%%%%%%%%%%%%%%%%%%%%%%%%%%%%%%%%%%%%%%%%%%%%%%%%%%%%%%%%%%%%%%%%%%%%%%%%%%%%%%%
%%%%%%%%%%%%%%%%%%%%%%%%%%%%%%%%%%%%%%%%%%%%%%%%%%%%%%%%%%%%%%%%%%%%%%%%%%%%%%%%%%%%%%%%%%%%%%%%%%%%%%
%%%%%%%%%%%%%%%%%%%%%%%%%%%%%%%%%%%%%%%%%%%%%%%%%%%%%%%%%%%%%%%%%%%%%%%%%%%%%%%%%%%%%%%%%%%%%%%%%%%%%%
\subsection{From the long-junction to short-junction limit: the scattering approach}
\label{shortjunctionsec}

The method described in the previous Section, which is based on the boundary conditions (\ref{eq:cond1}) - (\ref{eq:cond4}) and on the folded Hamiltonian,  is  valid  only in the limit in which $\Delta$ is the largest energy scale  (long-junction limit). We now wish to extend our investigation of the current also to the regime in which the junction length $L$ 
is much shorter than the superconducting correlation length $\xis$ [see Eq.~\eqref{longj}]. A different approach is thus necessary.  
To this purpose, we observe that since the Josephson current is an equilibrium current it can be computed also from the relation 
\begin{equation}
\label{eq:curr}
\langle I_\alpha \rangle = \frac{2e}{\hbar}\frac{ \partial \mathcal{F}_{\text{J}}}{ \partial \varphi_{\alpha}}~,
\end{equation}
where $\mathcal{F}_{\text{J}}$ is  the phase dependent  term  of the free energy, and~$\varphi_\alpha$ is the phase bias applied to layer $\alpha$.   The computation of the current through Eq.~(\ref{eq:curr}) offers the advantage that $\mathcal{F}_{\text{J}}$ can be evaluated from the knowledge of the eigenvalues of the system only (the eigenfunctions are not needed~\cite{beenakker_vanHouten_1992}).
Namely, one can write
\begin{equation}
\label{eq:josephson_free_energy}
 	\mathcal{F}_{\text{J}} = -\int\limits_0^{+\infty}d\eg\, \eg\rho(\eg)\,,
\end{equation}
where $\rho(\eg)$ is the density of states of the system and $\eg$ denotes the energy measured  from the Fermi energy.
The calculation of the free energy $\mathcal{F}_{\text{J}}$ for the system of two Josephson junctions coupled by an EC (Fig.~\ref{setup}) 
is one of the main results of the present Article. This computation enables us to recover the long-junction-limit result presented in Sec.~\ref{longjunctionsec} 
{\it via} an independent method and, most importantly, to obtain an expression for the current also in the short-junction limit.

Since the spectrum of the Hamiltonian in Eq.~\eqref{eq:ham} is symmetric around zero one can compute the energy density between zero and $+\infty$. 
Only the phase-dependent part of the density of states is relevant in the calculation of the Josephson current. The density of states $\rho$ depends 
on each of the four superconducting phases ($\varphi_{\alpha,L}$ and $\varphi_{\alpha,R}$) and it can be derived in a very simple and elegant way (see Appendix~\ref{app:scatt}) from 
the scattering matrix $S$ of the junction as follows~\cite{beenakker_1991, auerbach_1991}
\begin{equation}
\label{eq:scatt}
 	\rho(\eg) = \frac{1}{2\pi i}\frac{\partial}{\partial \eg}\ln (\det S) \,.
\end{equation}

For the sake of comparison, it is useful to recall what happens in a S-N-S junction (in the absence of an EC). The short-junction regime (where $\Delta$ 
is the smallest energy scale) is by far the simplest to treat, since the only phase-dependent feature of the spectrum is a single bound state with 
energy $\eg < \Delta$. In the long-junction limit~\cite{ishii_1970, bardeen_johnson_1972, kulik_1970} the number of bound states increases 
linearly with $L$ and, moreover, in order to properly evaluate the current, one has to take into account the continuum (for $\eg > \Delta$).
The new energy scale $|\Gamma|$, present in our system, enriches this picture, because one expects a contribution to the free energy due to 
the counterflow current, which is related to features of the spectrum at an energy $\eg \approx |\Gamma|$. Thus, even in the short-junction case, 
one has to compute $\rho(\varepsilon)$ at least up to $|\Gamma|$.

\subsubsection{Scattering matrix and density of states}

One possible approach to evaluate the scattering matrix is to generalize the method discussed in Ref.~\onlinecite{beenakker_1991} to the four terminal case. 
Here we propose, instead, a different approach, based on the following idea.
The superconducting contacts are not infinitely extended on the left and on the right, but truncated to a length $M$ so that the superconducting pairing potentials read~\cite{nota-on-L}
\begin{gather}
 \Delta_{T}(x) = \begin{cases}
			  0, & x\leq -M-L/2\\
                          \Delta e^{i\varphi_{T,L}}, & -M-L/2< x< -L/2\\
			  0 & -L/2, \leq x\leq L/2\\
			  \Delta e^{i\varphi_{T,R}}, & L/2<x<L/2+M\\
			  0, & x \geq L/2+M
                         \end{cases}
\end{gather}
and
\begin{gather}
\Delta_{B}(x) = \begin{cases}
		 0, & x \leq -M-L/2\\ 
                 \Delta e^{i\varphi_{B,L}}, & -M-L/2 < x <-L/2\\
		 0, & -L/2\leq x\leq L/2\\
		 \Delta e^{i\varphi_{B,R}}, &  L/2<x<L/2+M\\
		 0, & x \geq L/2+M
                \end{cases}~.
\end{gather}
Free-electron plane waves, present in the regions $x \leq -M-L/2$ and $x \geq L/2+M$, can therefore be used to define the total, block diagonal, scattering matrix
\begin{equation}
S=\left(
\begin{array}{cc}
S^+ & 0\\0 & S^-
\end{array}
\right)~,
\end{equation}
where $S^+$ and $S^-$ are defined through
\begin{equation}
\left(\begin{array}{c}
\psi_{T\uparrow +}(d) \\ \psi_{B\uparrow +}(-d) \\ \psi_{T\downarrow -}(-d) \\ \psi_{B\downarrow -}(d)
\end{array}\right)=
S^+ \left(
\begin{array}{c}
\psi_{T\uparrow +}(-d) \\ \psi_{B\uparrow +}(d) \\ \psi_{T\downarrow -}(d) \\ \psi_{B\downarrow -}(-d)
\end{array}
\right)~,
\end{equation}
and 
\begin{equation}
\left(\begin{array}{c}
\psi_{T\uparrow -}(-d) \\ \psi_{B\uparrow -}(d) \\ \psi_{T\downarrow +}(d) \\ \psi_{B\downarrow +}(-d)
\end{array}\right)=
S^- \left(
\begin{array}{c}
\psi_{T\uparrow -}(d) \\ \psi_{B\uparrow -}(-d) \\ \psi_{T\downarrow +}(-d) \\ \psi_{B\downarrow +}(d)
\end{array}
\right)\,.
\end{equation}
with $d = L/2+M$. Here, $\psi_{T\sigma+}(x)$ and $\psi_{B\sigma-}(x)$ denote the wave functions of a free ($\Delta=0$ and $\Gamma=0$) right-moving excitation, 
while $\psi_{T\sigma-}(x)$ and $\psi_{B\sigma+}(x)$ the left moving ones.

This truncation procedure offers the advantage that one can treat the discrete and continuous spectrum on the same footing, since now there are available free propagating channels also with energy $0<\eg<\Delta$, 
making it possible to define a scattering matrix in any energy range.
The exact energy density with infinitely extended superconducting contacts is found by taking the limit $M\to +\infty$ in Eq.~(\ref{eq:scatt}).
We have also checked that this truncation approach reproduces the well-known case of a standard ballistic two-terminal S-N-S junction.

A derivation of the relation~\eqref{eq:scatt} between the determinant of the scattering matrix and the density-of-states can be found in Appendix~\ref{app:scatt}, while in Appendix~\ref{app:det}  we show that the determinant of the scattering matrix appearing in Eq.~(\ref{eq:scatt}) can be expressed in terms of the transfer matrix $T$, which is easy to calculate for the present case of piecewise-constant potentials. 
In conclusion, through Eq.~(\ref{eq:josephson_free_energy}), the free energy can be computed numerically.
In the following subsections we will focus on two relevant limits where analytical results can be found.

\subsubsection{Phase-dependent contribution to the free energy}
\label{results}

We first restrict ourselves to the situation of exciton coupling characterized by the inequality $|\Gamma| \gg \Delta, E_{\text{T}}$, where 
$E_{\text{T}}=\hbar\vf/L$ is the Thouless energy.
Since in this regime the length $L$ of the bilayer is much greater than the EC coherence length, quasiparticle propagation in the bilayer is completely suppressed.
As a result, the density-of-states $\rho$ can be approximated by the sum of two contributions
\begin{equation}\label{eq:approx_split}
 \rho(\eg,\Delta,|\Gamma|) \simeq \rho_{\text{\tiny EC}}(\eg) + \rho_{\text{\tiny BS}}(\eg)~.
\end{equation}
The first term $\rho_{\text{\tiny EC}}(\eg)=\rho(\eg,0,|\Gamma|)$ accounts for the EC alone, while the second term $\rho_{\text{\tiny BS}}(\eg)=\rho(\eg,\Delta,\infty)$ is related to the superconducting electrodes coupled to an EC characterized by an infinite gap.

Combining Eq.~(\ref{eq:approx_split}) with  Eq.~(\ref{eq:josephson_free_energy}), the free energy can be  written as the sum of  two contributions
\begin{equation}\label{eq:splitFj}
\mathcal{F}_{\text{J}} \simeq \mathcal{F}_{\text{\tiny EC}} + \mathcal{F}_{\text{\tiny BS}}~.
\end{equation}
The first term, in the limit of large 
$|\Gamma|$, turns out to be (see Appendix~\ref{app:cond_curr})
\begin{equation}\label{eq:ex_contrib}
\mathcal{F}_{\text{\tiny EC}} = \frac{\hbar\vf}{2\pi L}(qL)^2 + O(|\Gamma|^{-2})~,
\end{equation}
which describes an excitonic supercurrent contribution, encoded in the phase winding $q$ of the EC order parameter.

The second term in Eq.~(\ref{eq:splitFj})  can be understood as the contribution to the free energy due to bound states. 
Indeed,  in the limit $M\to+\infty$, $\rho_{\text{\tiny BS}}(\eg)$ consists of two $\delta$-functions (see Appendix~\ref{app:bound}), {\it i.e.} 
\begin{equation}
\rho_{\text{\tiny BS}}(\eg)=\delta (\eg-\eg_{L})+\delta (\eg-\eg_{R}) \,,
\end{equation}
where
\begin{equation}\label{eq:bound_l}
\eg_{L} = \Delta\left|\cos\left[\frac{\varphi_L}{2}-\gamma+\frac{qL}{2}\right]\right|
\end{equation}
and
\begin{equation}\label{eq:bound_r}
\eg_{R}=\Delta\left|\cos\left[\frac{\varphi_R}{2}-\gamma-\frac{qL}{2}\right]\right|~.
\end{equation}
Here
\begin{equation} \label{varphiL-def}
\varphi_L = \varphi_{T,L}-\varphi_{B,L}
\end{equation}
and
\begin{equation}\label{varphiR-def}
\varphi_R = \varphi_{T,R}-\varphi_{B,R} 
\end{equation}
denote the phase differences between left and right electrodes, respectively.

The wave functions of the bound states are exponentially decaying both in the EC and in the superconducting contacts.
In particular, they are localized at the superconductor - EC interfaces and they exist independently one of each other.
As a result one obtains the total free energy of the system
\begin{eqnarray}\label{eq:main}
\mathcal{F}_{\text{J}}(qL, \gamma, \varphi_L, \varphi_R) &=& \frac{\hbar\vf}{2\pi L}(qL)^2 -\Delta\left|\cos\left[\frac{\varphi_L}{2}-\gamma+\frac{qL}{2}\right]\right| \nonumber \\
&-&\Delta\left|\cos\left[\frac{\varphi_R}{2}-\gamma-\frac{qL}{2}\right]\right|~.
\end{eqnarray}
This is the main result of this Section. 

We now proceed to minimize the total free energy $\mathcal{F}_{\text{J}}(qL, \gamma,  \varphi_L, \varphi_R)$ with respect to $\gamma$. We find that the optimal value of $\gamma$ has to be of the form
\begin{equation}\label{eq:optimalgamma}
\gamma = \frac{\varphi_L + \varphi_R}{4}  + n \frac{\pi}{2}~
\end{equation}
with $n$ a relative integer. This optimal value of $\gamma$ physically ensures that the currents flowing in the external leads have equal value but opposite sign on the left and on the right. 
When $\varphi_{L}+\varphi_{R}$ is altered, only the overall phase $\gamma$ of the EC responds. 

Substituting Eq.~(\ref{eq:optimalgamma}) in Eq.~(\ref{eq:main}) we find an expression for the free energy as a function of $qL$ only:
\begin{eqnarray}\label{eq:optimalgammafreeenergy}
{\widetilde {\cal F}}_{\rm J}(qL) &\equiv& \min_{\gamma}[\mathcal{F}_{\text{J}}(qL, \gamma)] = \frac{\hbar\vf}{2\pi L}(qL)^2 \nonumber \\
&-& 2\Delta\cos\left(\frac{\varphi_R-\varphi_L-2qL}{4}\right)~,
\end{eqnarray}
which holds when $\varphi_R - \varphi_L - 2qL \in \left[\,-\pi;\;\pi\,\right]$. For values of $\varphi_R - \varphi_L - 2 qL$ outside this interval one takes advantage of the $2\pi$-periodicity. 

We clearly see from Eq.~(\ref{eq:optimalgammafreeenergy}) that the free energy (and thus the associated Josephson current) depends only on phase difference $\varphi_R-\varphi_L$, 
which, by using Eqs.~(\ref{varphiT-def})-(\ref{varphiB-def}) and~(\ref{varphiR-def})-(\ref{varphiL-def}), can also be written as
\begin{equation} \label{phiLR-TB}
\begin{split}
\varphi_R-\varphi_L  = \varphi_T-\varphi_B~.
\end{split}
\end{equation}
Minimizing ${\widetilde {\cal F}}_{\rm J}(qL)$ with respect to $qL$ enforces the condition that the EC counterflow current matches the supercurrents carried by the condensate in the superconducting electrodes. In what follows we will drop the tilde symbol over ${\cal F}_{\rm J}$ for notational simplicity.

\begin{figure}
\includegraphics{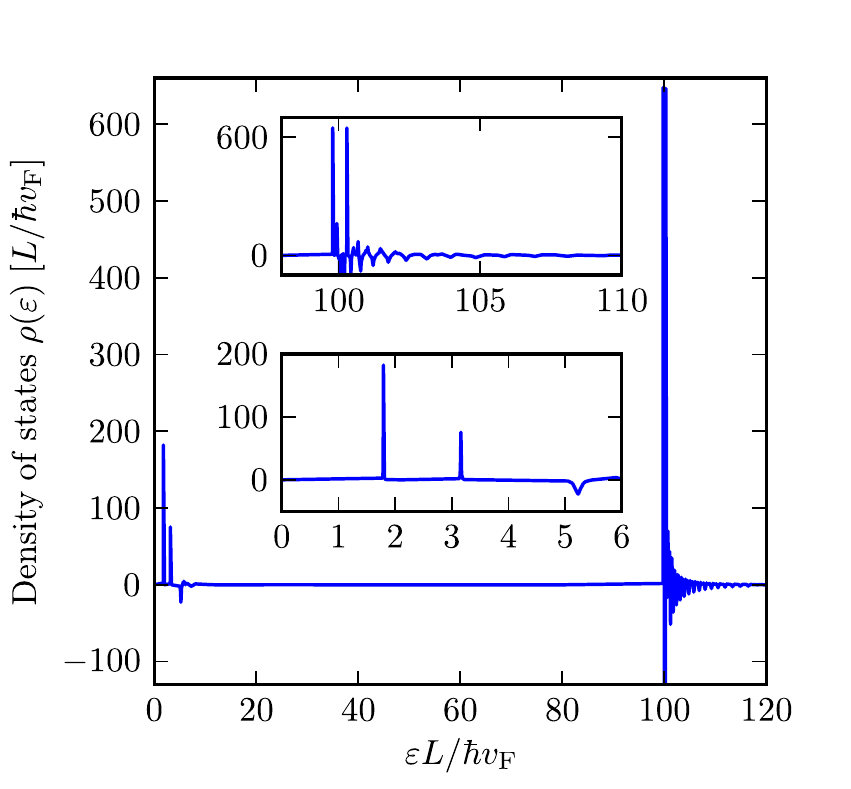}
\caption{\label{DOS} (Color online) In the main panel we show the density-of-states of the system $\rho(\eg,\Delta,|\Gamma|)$ (in units of $1/E_{\rm T}$) as a function of energy $\eg$ (in units of  $E_{\rm T}$) 
as computed numerically from Eq.~\eqref{eq:scatt} for a random choice of the phases $\varphi_i,\gamma$, and $qL$ and for $\Delta=5~E_{\rm T}$ and $|\Gamma|=100~E_{\rm T}$. Lower inset: the contribution due to the presence of the superconducting electrodes $\rho_{\text{\tiny BS}}(\eg)$ as a function of $\eg$. Upper inset: the contribution due to the EC $\rho_{\text{\tiny EC}}(\eg)$ as a function of $\eg$.  These two contributions can be considered as independent in the strong coupling limit $|\Gamma| \gg \Delta,\hbar \vf /L$.} 
\end{figure}

In the remaining parts of this Section we will  use Eq.~(\ref{eq:optimalgammafreeenergy}) to derive the Josephson current in the long- and short-junction regimes. 
Before doing that, we wish to mention that we have checked numerically the validity of Eq.~(\ref{eq:approx_split}) by comparing the three quantities 
$\rho(\eg,\Delta,|\Gamma|)$, $\rho_{\text{\tiny EC}}(\eg)$, and $\rho_{\text{\tiny BS}}(\eg)$. The result of this comparison is reported in Fig.~\ref{DOS}, 
where we plot these three quantities versus $\eg$. The bound-state contribution to the density-of-states $\rho_{\text{\tiny BS}}(\eg)$, plotted in the lower inset, mainly consists of two sharp peaks 
and is negligible for $\eg >10~E_{\text{T}}$ (the small downward peak is related to the continuum spectrum at energies $\eg>\Delta$ and disappears in the the limit $M\to +\infty$). 
The EC contribution to the density-of-states $\rho_{\text{\tiny EC}}(\eg)$, shown in the upper inset, is essentially zero up to $\eg \sim|\Gamma|$.
It turns out that the density-of-states $\rho(\eg,\Delta,|\Gamma|)$ evaluated numerically from Eq.~(\ref{eq:scatt}), and shown in the main panel, coincides with the sum of the data reported in the insets.
\subsubsection{Long-junction limit ($L \gg \xis$)}
\label{resultslong}
The current in the long-junction regime has already been investigated in Ref.~\onlinecite{dolcini_2009} with the boundary conditions and folding method, as  outlined in Sec.~\ref{longjunctionsec}. Here we provide an alternative derivation of Eq.~(\ref{eq:to}) based on the truncation method, introduced at the beginning of this Section, and on Eq.~\eqref{eq:main}. This proves the soundness of the method.

Since the long-junction regime is characterized by $\Delta \gg \hbar\vf/L$, one can, to a first approximation,  neglect the quadratic term in the phase winding $q$ in the analytical expression for the free energy, {\it i.e.} the first term in the r.h.s. of Eq.~(\ref{eq:optimalgammafreeenergy}). Minimization of the second term in the r.h.s. of Eq.~(\ref{eq:optimalgammafreeenergy}) with respect to $qL$ yields 
\begin{equation}\label{eq:phase_wind}
qL = \frac{\varphi_R-\varphi_L}{2} + n\pi~,
\end{equation}
with $n$ a relative integer. This is the same result found in Eq.~\eqref{eq:qll}. There it was derived as a condition for the existence of solutions of the Hamiltonian, while here we interpret 
it as a relation enforced by the EABS, which is increasingly better satisfied as the ratio $\Delta/E_{\rm T}$ grows.

Eqs.~(\ref{eq:optimalgamma})-(\ref{eq:phase_wind}) can be interpreted as ``\emph{phase anchoring conditions}" imposed by the presence of EABS, since the phase of the condensate at both interfaces and the 
superconducting phases are no longer independent.

The integer $n$ in Eq.~(\ref{eq:phase_wind}) is fixed in such a way to minimize the first term of the free energy in Eq.~\eqref{eq:optimalgammafreeenergy}. The latter thus reads
\begin{equation}
 \mathcal{F}_{\text{J}} = \frac{\pi\hbar\vf}{2L}\left[\frac{\varphi_R-\varphi_L}{2\pi}
-\mathrm{nint}\left(\frac{\varphi_R-\varphi_L}{2\pi}\right)\right]^2\,
\end{equation}
where $\mathrm{nint}(x)$ denotes the integer closest to $x$, or
\begin{equation}\label{eq:free2}
\mathcal{F}_{\text{J}}= \frac{\hbar\vf}{2\pi L}\left(\frac{\varphi_R-\varphi_L}{2}\right)^2~,
\end{equation}
if the phase difference $\varphi_R-\varphi_L$ is restricted to vary in the interval $\left[\,-\pi;\;\pi\,\right]$.

Since an EABS decays exponentially in the bilayer, current can flow only through the EC and can be calculated as following:
\begin{equation}\label{eq:ex_curr}
 I_{T/B}(q) = \pm\frac{e}{\hbar}\frac{d\mathcal{F}_{\text{\tiny EC}}}{d(qL)} = \alpha\frac{e\vf}{\pi L}(qL)~.
\end{equation}
As a result, the Josephson current in a given layer is obtained by
inserting Eq.~\eqref{eq:phase_wind} into Eq.~\eqref{eq:ex_curr} to obtain
\begin{equation}\label{eq:curr2}
I_{T/B}(\varphi_R-\varphi_L) = \pm\frac{e\vf}{L} \left(\frac{\varphi_R-\varphi_L}{2\pi} \right)\,,
\end{equation}
with $\varphi_R-\varphi_L \in \left[\,-\pi;\;\pi\,\right]$. Eq.~(\ref{eq:curr2}) can also be derived using Eq.~\eqref{eq:curr}.
For the top layer, for example, Eq.~\eqref{eq:curr} takes the form 
\begin{equation}
I_{T} = \frac{2e}{\hbar}\frac{\partial \mathcal{F}_{\text{J}}}{\partial \varphi_T}~.
\end{equation}
Using Eq.~(\ref{phiLR-TB}) one therefore obtains
\begin{equation}
I_{T} = \frac{e\vf}{L}\left(\frac{\varphi_T-\varphi_B}{2\pi}\right)~,
\end{equation}
which coincides with Eq.~\eqref{eq:curr2}.

Note that the current has exactly the same functional form of a standard long Josephson junction at $T = 0$ [see Eq.~\eqref{eq:long}] albeit with a critical current reduced by a factor of two.
The significance of the particular combination of phases appearing in the argument of the current will be discussed below.

\subsubsection{Short-junction limit ($ L \ll \xis$)}
\label{sec:ballistic-short}
In the short-junction regime the dominant term in the free energy is the EC counterflow current contribution
\begin{equation}
\mathcal{F}_{\text{\tiny EC}} = \frac{\hbar\vf}{2\pi L}(qL)^2 = \frac{\Delta}{\epsilon}(qL)^2~,
\end{equation}
where $\epsilon \equiv 2L/\xis \ll 1$.  Minimizing Eq.~(\ref{eq:optimalgammafreeenergy}) with respect to $qL$ one gets
\begin{eqnarray}\label{eq:winding}
 qL &=& \frac{\epsilon}{2}\sin\left(\frac{\varphi_R-\varphi_L}{4}-\frac{qL}{2}\right) \nonumber \\
 &\approx& \frac{\epsilon}{2}\sin\left(\frac{\varphi_R-\varphi_L}{4}\right) + {\cal O}(\epsilon^2)~,
\end{eqnarray}
where $qL$ inside the phase argument can be neglected since it is a small quantity of order $\epsilon$ [this can be easily seen from  Eq.~\eqref{eq:winding} itself]. 
Eq.~(\ref{eq:winding}) can be interpreted as the equivalent of Eq.~(\ref{eq:phase_wind}) in the short-junction limit.

Substituting Eq.~(\ref{eq:winding}) in Eq.~\eqref{eq:optimalgammafreeenergy} and neglecting terms of order $\epsilon$ yields
\begin{equation}\label{eq:free4}
 \mathcal{F}_{\text{J}} = -2\Delta\cos\left(\frac{\varphi_R-\varphi_L}{4}\right)~.
\end{equation}
Now, as before, the current can be calculated by either inserting Eq.~\eqref{eq:winding} in Eq.~\eqref{eq:ex_curr} or applying Eq.~\eqref{eq:curr} 
to Eq.~\eqref{eq:free4} (differentiating with respect to $\varphi_T$ for the current on the top layer and with respect to $\varphi_B$ for the current in the bottom layer).
In both cases the result is
\begin{equation}\label{eq:short_current}
I_{\text{T/B}}(\varphi_T-\varphi_B) = \pm\frac{e\Delta}{\hbar}\sin\left(\frac{\varphi_T-\varphi_B}{4}\right)~,
\end{equation}
with $\varphi_T-\varphi_B \in \left[\,-\pi;\;\pi\,\right]$.
\begin{figure}
 \includegraphics{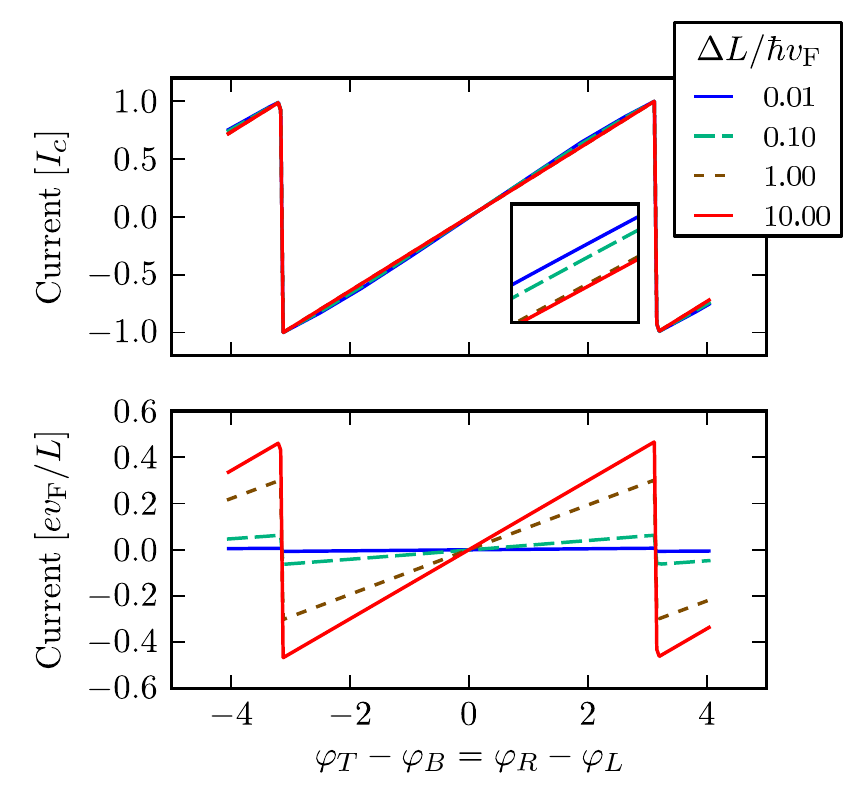}
\caption{\label{currf}(Color online) The Josephson current as a function of $\varphi_T-\varphi_B$ normalized in two different ways. In the upper panel the current 
is normalized with respect to the critical current, while in the lower panel it is expressed in units of $e\vf/L$.}
\label{figure_current}
\end{figure}
\subsubsection{Crossover regime ($L \sim \xis$)}
In the intermediate or crossover regime ($L \sim \xis$), the Josephson current can be understood as interpolating 
between long- and short-junction limits.  In this regime a derivation of analytical expressions is particularly difficult.  A numerical analysis leads to 
the results summarized in Fig.~\ref{figure_current}. It is straightforward  to minimize Eq.~(\ref{eq:optimalgammafreeenergy}) 
with respect to $qL$, the parameter describing the phase winding in the bulk of the EC. As we have seen above in Eq.~(\ref{eq:optimalgammafreeenergy}), 
after minimization, the final result depends only on the phase difference $\varphi_R-\varphi_L = \varphi_T-\varphi_B$. 
This can be understood by noting that equal shifts in $\varphi_L$ and $\varphi_R$ can always be reabsorbed
in $\gamma$, keeping the currents flowing in the left and right leads constant. 
Thus the only relevant degree-of-freedom left is $\varphi_R-\varphi_L$ and, even in the crossover regime, we have 
the same behavior found previously (namely exciton blockade and superdrag) albeit with a different 
current-phase relationship. As shown in Fig.~\ref{figure_current}, the current-phase relation does not change dramatically
during the evolution from the short- to the long-junction limit. When $I_{\text{T}}$ is normalized with respect to the critical current (upper panel), the various curves differ by a small amount when $\Delta$ changes by three orders of magnitude (see inset in the upper panel of  Fig.~\ref{figure_current}). Nevertheless, it is important to study the peculiarities of the current-phase relation since it can reveal a wealth 
of information about the microscopic processes that influence the supercurrent~\cite{chialvo_2010}.

A change in the current-phase relation is also expected when the magnitude of the superconducting gaps in the four electrodes is different (here we assumed for simplicity that they are all equal to $\Delta$). The energy scale in 
$\epsilon_R$ and $\epsilon_L$ [Eqs.~(\ref{eq:bound_l}) and~(\ref{eq:bound_r})], and as a consequence the 
supercurrent, will change.  Again these will be only quantitative effects, the fact that the supercurrent depends solely on 
$\varphi_R - \varphi_L$ is unaltered.

\section{Josephson current in the tunneling regime}
\label{sec:tunneling}

So far, we have analyzed the behavior of the current in various regimes of the junction length, assuming  {\it perfectly transparent} contacts between the bilayer EC and the superconducting electrodes. 
We will now analyze the opposite regime, in which the contacts to the superconducting electrodes have low transparencies. In this case we employ a tunneling Hamiltonian defined by
\begin{equation}
\mathcal{H} = \sum_{i = L, R} \sum_{\alpha = T, B} \mathcal{H}_{\text{\tiny S}, i,\alpha} + \mathcal{H}_{\text{\tiny EC}}
+ \mathcal{H}_{\text{T}}~,
\end{equation}
with the tunneling coupling given by
\begin{equation}
\mathcal{H}_{\text{T}} = \sum_{i,\alpha,\sigma}\int d{\bm x} d{\bm x}' 
t_{i,\alpha}({\bm x}, {\bm x'}) \psi^\dagger_{\text{\tiny S},\sigma,i,\alpha}({\bm x})
\psi_{\text{\tiny EC},\sigma,\alpha}({\bm x'}) + \mathrm{H.c.}
\end{equation}
while the superconductors and exciton condensate  Hamiltonians are expressed in the  mean-field approximation with order 
parameters $\Delta = |\Delta|e^{i\varphi_{i,\alpha}}$ and $\Gamma(x) = |\Gamma|e^{i(\gamma+qx)}$.

The pertubative expansion in powers of the tunneling matrix element $T_{i,\alpha}({\bm x}, {\bm x'})$ can be set as follows. 
Let ${\cal K}={\cal K}_0 + {\cal K}_1$ where ${\cal K}_0 = \sum_{i,\alpha} \mathcal{H}_{\text{S},i,\alpha} + \mathcal{H}_{\text{EC}}-\mu N$ 
and ${\cal K}_1={\cal H}_{\rm T}$. One has  to calculate the {\it phase-dependent} term in the grand-potential 
\begin{equation}
\Omega = -\frac{1}{\beta}~\ln{{\cal Z}}~,
\end{equation}
where
\begin{equation}
{\cal Z}={\rm Tr}[e^{-\beta({\hat {\cal H}}-\mu {\hat N})}]
\end{equation}
is the partition function and $\beta=(\kb T)^{-1}$ is the inverse temperature.
In order to capture all the relevant phase-dependent terms one has to expand the free energy up to {\it fourth order} in the tunneling:
\begin{equation}
\Omega = \Omega_0 - \frac{1}{\beta}\left\{\frac{1}{2!}\langle {\cal O}_2\rangle_0+
\frac{1}{4!}\langle{\cal O}_4\rangle_0\right\}~.
\end{equation}
The mean $\left\langle\dots\right\rangle_0$ is defined as a mean on the thermal ensemble
$\left\langle\mathcal{O}\right\rangle_0 \equiv {\rm Tr}(e^{-\beta {\cal K}_0} {\cal O})/{\cal Z}_0 \equiv {\rm Tr}(\rho_0 {\cal O})$ 
and $\mathcal{O}_2$ and $\mathcal{O}_4$ are imaginary-time ordered integrals
\begin{equation}
\mathcal{O}_2 = \left[\int_0^\beta d\tau\, \int_0^\beta d\tau'\,\mathrm{T}_\tau \left[\mathcal{H}_\text{T}(\tau)\mathcal{H}_\text{T}(\tau')\right]\right]_{\text{ctd}}~,
\end{equation}
and
\begin{equation}
\mathcal{O}_4 = \left[\int_0^\beta d\tau_1\dots \int_0^\beta
d\tau_4~\mathrm{T}_\tau\left[\mathcal{H}_\text{T}(\tau_1)\dots
\mathcal{H}_\text{T}(\tau_4)\right]\right]_{\text{ctd}}~.
\end{equation}
By $[\dots]_{\text{ctd}}$ we mean that only fully-connected contractions are included.
In the following we will focus only on the phase-dependent contributions. 

The zeroth-order term $\Omega_0$ gives 
the  energy cost due to the   phase winding $q$ of the EC order parameter
\begin{equation}
\mathcal{F}_{\text{\tiny EC}} = \frac{\hbar \vf}{2\pi L}(qL)^2~.
\end{equation}
The contributions coming from the superconducting contacts necessarily stem from the 
fourth-order term since the superconductors are coupled only through the condensate. The phase-dependent 
ones involve the anomalous propagators both for the superconductors and the EC:
\begin{multline}\label{eq:op}
\frac{1}{24}\left\langle\mathcal{O}_{4,\text{phase-dependent}}\right\rangle_0 
= \frac{1}{24}\int_0^\beta d\tau_1\,\dots\int_{0}^\beta d\tau_4\times\\ \,\langle \psi^\dagger_{\text{\tiny S},\sigma_1,i_1,\alpha_1}(\tau_1)
\psi^\dagger_{\text{\tiny S},\sigma_2,i_2,\alpha_2}(\tau_2)\rangle_0\times \\
\langle \psi_{\text{\tiny EC},\sigma_1,\alpha_1}(\tau_1)
\psi^\dagger_{\text{\tiny EC},\sigma_3,\alpha_3}(\tau_3)\rangle_0 \times \\
\langle \psi_{\text{\tiny EC},\sigma_2,\alpha_2}(\tau_2) 
\psi^\dagger_{\text{\tiny EC},\sigma_4,\alpha_4}(\tau_4)\rangle_0 \times \\
\langle\psi_{\text{\tiny S},\sigma_4,i_4,\alpha_4}(\tau_4)
 \psi_{\text{\tiny S},\sigma_3,i_3,\alpha_3}(\tau_3)
\rangle_0 + \text{H.c.}
\end{multline}
where
\begin{multline}
\langle \psi^\dagger_{\text{\tiny S},\downarrow,L,T}(\tau_1,{\bm x}_1)
\psi^\dagger_{\text{\tiny S},\uparrow,L,T}(\tau_2,{\bm x}_2)\rangle_0 = \\
-\frac{1}{\beta}\sum_n e^{-i\omega_n(\tau_1-\tau_2)}
\frac{1}{S} \sum_{\bm p} e^{i{\bm p}({\bm x}_1-{\bm x}_2)}
\frac{\Delta^*_{L,T}}{\omega_n^2+\xi_{\bm p}^2+|\Delta_{L,T}|^2}
\end{multline}
and very similar expressions hold for the other contacts ($\xi_{\bm k} = \hbar^2{\bm k}^2/2m-\mu$). We take the magnitude of the parameters $|\Delta_{i,\alpha}|$ equal for all the contacts. Only the phases vary.
For the EC ``interlayer propagators" we find
\begin{multline}
\langle \psi_{\text{\tiny EC},\sigma,T}(\tau_1,{\bm x}_1)
\psi^\dagger_{\text{\tiny EC},\sigma,B}(\tau_3,{\bm x}_3)\rangle_0 = \\
-\frac{1}{\beta}\sum_n e^{-i\omega_n(\tau_1-\tau_3)}
\frac{1}{S}\sum_{\bm p}e^{i{\bm p}({\bm x}_1-{\bm x}_3)}
\frac{\Gamma}{\omega_n^2+\xi_{\bm p}^2+|\Gamma|^2}~.
\end{multline}
The integrations on the tunneling amplitudes in Eq.~\eqref{eq:op} are left implicit for simplicity, although 
their calculation can be carried out analytically. Only proper combinations of the indices produce a relevant contribution. 
Cooper pairing requires $\sigma_1=-\sigma_2=-\sigma_3=\sigma_4$, and since we are 
interested in the coupling between different layers $\alpha_1=\alpha_2=T$ and
$\alpha_3=\alpha_4=B$. Finally one requires $i_1=i_2=i_3=i_4=L,R$.  Indeed in the other case
$i_1=i_2\neq i_3=i_4$ one has EC propagators connecting contacts on opposite sides of the junction, which are exponentially suppressed by the gap $|\Gamma|$, though. We thus have 
two different contributions coming from terms localized at the two sides. Reshuffling the indices is equivalent to a permutation of the integration variables, which cancels the $1/4!$ prefactor. 
The  free spin index gives an extra factor two. We take the phase of the EC order parameter as constant in the region of the left contacts even though it is supposed to wind along the junction, {\it i.e.} the contact regions are small. This means that $\Gamma = \Gamma(0) = |\Gamma|e^{i\gamma}$. Another approximation will be to take $t_{i,\alpha}({\bm x},{\bm x}') = t\delta^{(2)}({\bm x}-{\bm x}')$ in the contact region. 
This leads to the result
\begin{equation} 
\mathcal{F}_L=-\frac{4}{\beta}|t|^4|\Delta|^2|\Gamma|^2 I(\beta,|\Delta|,|\Gamma|)~\cos(2\gamma-\varphi_L)\,, \label{F_L}
\end{equation}
where 
\begin{multline}
I(\beta,|\Delta|,|\Gamma|) = \\
\sum_n\sum_{\bm p}\left(\frac{1}{\omega_n^2+\xi_{\bm p}^2+|\Delta|^2}
\frac{1}{\omega_n^2+\xi_{\bm p}^2+|\Gamma|^2}\right)^2~.
\end{multline}
This integral can be calculated, but its exact value is not of importance for the purpose of the present Article. 
An analogous calculation for the right contact produces
\begin{multline}
\mathcal{F}_R=-\frac{4}{\beta}|t|^4|\Delta|^2|\Gamma|^2 I(\beta,|\Delta|,|\Gamma|)\times \\
\cos(2\gamma+2qL-\varphi_R)\,. \label{F_R}
\end{multline}
Note that in the left-contact contribution [Eq.~(\ref{F_L})]  the value $2 \gamma$ of the EC phase   
at left side of the junction appears, while in the right-contact contribution  [Eq.~(\ref{F_R})] the value 
$2 \gamma+2qL$ at the right side appears.  These phases are coherently coupled with the difference 
of the left and right superconducting phases $\varphi_L$ and $\varphi_R$, respectively. Together with the EC bulk contribution this 
produces a free energy, which is in form very similar to the one studied in the case of ideal interfaces, {\it modulo} a different 
functional dependence on phase [$\cos(\varphi)$, typical of a tunnel junction, instead of $\cos(\varphi/2)$].  
As we said earlier, the value of the prefactor is not so relevant here since the main result that can be derived from the total free-energy functional,
\begin{equation}
\begin{split}
\mathcal{F} = \mathcal{F}_{\text{\tiny EC}}(qL)+
\mathcal{F}_{L}[2\gamma-(\varphi_{L,T}-\varphi_{L,B})]\\
+\mathcal{F}_{R}[2\gamma+2qL-(\varphi_{R,T}-\varphi_{R,B})]~,
\end{split}
\end{equation}
is that the Josephson current is of the form
\begin{equation}
I_{T/B} = \pm~I_c\sin\left(\frac{\varphi_T-\varphi_B}{2}\right)~,
\end{equation}
for $\varphi_T-\varphi_B\in \left[\,-\pi;\;\pi\,\right]$ and then extended by periodicity. 
The derivation is very similar to the one given for the ballistic short junction  (Sec.~\ref{sec:ballistic-short}).

\section{Topologically protected qubits}
\label{protected}

\begin{figure}[th]
\includegraphics[width=3in]{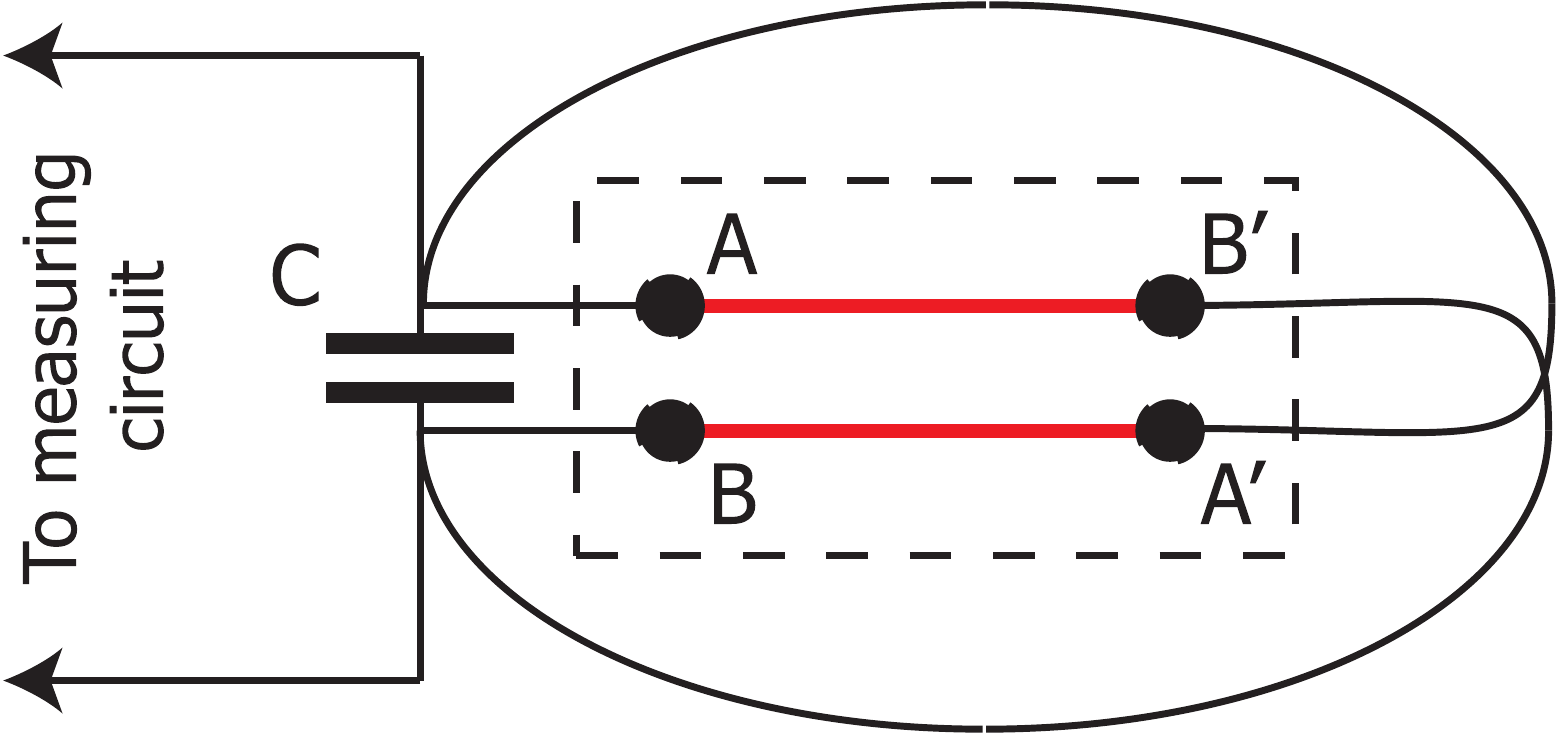}
\caption{\label{circuit}A four terminal device studied which realizes a 
protected qubit. The connections between pairs of terminals
$A,A'$ and $B,B'$ ensure that the phase differences $\varphi_{T}=\varphi(B')-\varphi(A)$
and $\varphi_{B}=\varphi(A')-\varphi(B)$ are opposite: $\varphi_{B}=-\varphi_{T}$.
This results in a doubly periodic free energy dependence on the phase
difference $\phi=\varphi(B)-\varphi(A)$ and formation of two degenerate
states that can be used for quantum computation. }
\end{figure}

A very important consequence of the fact that the Josephson energy depends
only on the difference $\varphi_{T}-\varphi_{B}$ [see Eqs.~(\ref{eq:to}) and~(\ref{eq:short_current}) for long and short junctions, respectively] is the appearance
of an almost \emph{exact} double periodicity in the energy of the circuit illustrated in Fig.~\ref{circuit}.  
The double periodicity is  analogous to the one suggested originally by Kitaev~\cite{kitaev} for Josephson junction arrays with
a similar property, {\em i.e.} energy dependence on the variable $\varphi_{T}-\varphi_{B}$ only. The cross connections in Fig.~\ref{circuit} ensure that
fluctuations which change $\phi_T-\phi_B$ are suppressed,
leaving $\phi=\varphi_{T} = - \varphi_{B}$ as the only degree of freedom.
An energy that is function of the phase difference $\varphi_{T}-\varphi_{B}$
may be expressed as $F(\varphi_{T}-\varphi_{B})=F(2\phi)$ which is doubly periodic
compared to a usual Josephson energy. 

Exact double periodicity of the free energy is quite generally 
promising for decoherence-free quantum computation\cite{Ioffe2002}.
It implies that the energy of the resulting circuit has two distinguishable
minima located at $\phi=0$ and $\phi=\pi$ that are separated by
a maximum at $\phi=\pm\pi/2$ with energy $E_{2}.$  The two quantum states
$|0 \rangle$ and $|\pi \rangle$, corresponding to phase differences $\phi=0$ and
$\phi=\pi$, can be used as the logical states of a quantum computation. The
fact that their energies are equal implies absence of the dephasing
processes. The large barrier between them implies that there is no tunneling and thus no decay. 

Decoherence-free quantum computation becomes possible only if
one can satisfy a number of constraints on the energy scales that
we discuss below. We now show that these constraints might be easier to
satisfy for the four-terminal device discussed in this Article than
for the original ``Josephson-current mirror" suggested by Kitaev. 

The height of the barrier is $E_{2}=\pi \hbar \vf/(2L)$ for a long one-dimensional contact. 
In general, for a more realistic two-dimensional system, {\it e.g.} a graphene sheet with transverse size $d$, we expect
$E_{2}=\gamma(\kf d) \hbar v_{F}/L$ where $\gamma\sim 1$ is  a numerical coefficient.
As we explain below, the optimal value of $E_{2}$ is in the range
$1-10~{\rm K}$, which is achieved for one-dimensional contacts of length
$L=1-10~\mu {\rm m}$. 

The  energy difference between the two minima of the potential is due
to single-electron tunneling through the contact which has been neglected up to now 
in the discussion because it is exponentially small: $\delta E\sim E_{2}\exp(-L/\xie)$.
Because $E_{2}$ decreases slowly with the increase of the length
of the contact and is expected to be proportional to the width of
the contact for two-dimensional structures it should be possible to
produce a circuit characterized by energy $E_{2}/\kb \gtrsim 1~{\rm K}$ and
very small splitting $\delta E/\kb \lesssim10^{-10}~{\rm K}$.
Almost exact double periodicity of the energy of the circuit implies
that the two quantum states, $|0 \rangle$ and $|\pi \rangle$, remain coherent for a
time that is limited by the shortest of the inverse splitting time $\hbar/\delta E\gtrsim  1~{\rm s}$
and the time required for quantum or thermal tunneling across the barrier.
The quantum tunneling is due to the charging energy, $E_{C}=e^{2}/2C$,
of the mesoscopic device. This energy can be made small by attaching
a capacitor as shown in Fig.~\ref{circuit} which makes the tunneling
amplitude $t\sim E_{J}^{3/4}E_{C}^{1/4}\exp\big[-(\pi/8)\sqrt{E_{2}/E_{C}}\,\big]$
of the same order as the splitting between the maxima.  For 
$E_{2}\gtrsim1{\rm K}$ the capacitor should be of  the order of $C\sim 1~{\rm pF}$.
Thermal tunneling across the barrier is exponentially small at
the typical base temperature of $T\sim 20~{\rm mK}$: $\tau_{th}\sim\omega_{p}^{-1}\exp(-E_{2}/\kb T)\sim 1~{\rm s}$
where $\omega_p$ is characteristic frequency of the quantum oscillations
within each minima that is due to charging energy of the device: $\hbar\omega_{p}=(4/\pi)\sqrt{E_{2}E_{C}}$.

Thus, a device of this type should preserve coherence between the
two quantum states for a time of the order of $1~{\rm s}$. Two quantum
states can be controlled by including additional SQUID loops in the
connecting loops that vary the energy of the logical states depending
on magnetic field, as discussed by Kitaev~\cite{kitaev}. 
The operations on the quantum states do not excite quasiparticles in the superconductors 
provided that the energies remain smaller than the superconducting
gap. This condition translates into the requirement that $E_{2}\lesssim\Delta$.
On the other hand one wants to keep $E_{2}$ as large as possible
to suppress thermal excitations. These two conditions together imply
that the range $E_{2}/k_B \sim 1-10~{\rm K}$ mentioned above is optimal. 

\section{Conclusions}
\label{conc}
In this Article we have calculated the Josephson current between two pairs of superconducting terminals 
coupled by a bilayer electron system that hosts an EC.
We have focused on the regime of strong exciton coupling where the bilayer gap $|\Gamma|$ 
is the largest energy scale. In this limit quasiparticles cannot propagate through the bilayer and the Josephson current is entirely due to the 
conversion of Cooper pair current into counterflow excitonic supercurrent.  
We have considered both the short- ($L \ll \xis$) and ($L\gg \xis$) long-junction regimes.
The Josephson current, at zero temperature, is given by Eq.~\eqref{eq:short_current} and Eq.~\eqref{eq:to} 
in the short- and long-junction regime, respectively. Results for the intermediate  regime are plotted in Fig.~\ref{figure_current}.
The Josephson current, calculated numerically, is plotted in Fig.~\ref{currf} as a function of the phase difference $\varphi_T-\varphi_B$ for different values 
of the ratio $\Delta/E_{\text{T}}$ spanning both regimes.

Surprisingly, these two results [Eqs.~\eqref{eq:short_current} and \eqref{eq:to}] can be simply obtained starting from the standard result for the Josephson current $I(\varphi)$ for the two separate (top and bottom layer) junctions and applying the substitution $\varphi\to (\varphi_T-\varphi_B)/2$. Namely, for a standard S-N-S ballistic short junction at $T=0$
\begin{equation}
 I(\varphi) = \frac{e\Delta}{\hbar}\sin\left(\frac{\varphi}{2}\right)~,
\end{equation}
whereas for a S-N-S long junction at $T=0$
\begin{equation}\label{eq:long}
 I(\varphi) = \frac{e\vf}{L}\frac{\varphi}{\pi}~.
\end{equation}
In particular, in the short-(long-) junction case the critical current is reduced by a factor $1/\sqrt{2}$ ($1/2$).

The reason why the superconducting  phases enter only the Josephson current expression in the combination $(\varphi_T-\varphi_B)/2$ 
may be traced back to the relation between the superconducting contact phases and the EC phase, which is energetically forced by the EABS [as an example, see the ``phase anchoring conditions" in Eqs.~(\ref{eq:optimalgamma})-(\ref{eq:phase_wind}) or~(\ref{eq:optimalgamma})-(\ref{eq:winding})]. From the derivation given in Appendix~\ref{app:bound} it emerges that EABS encode a process of correlated Andreev reflection, which is the only current conversion mechanism available in the presence of a strong exciton condensate. For each Cooper pair absorbed in the top layer by the condensate a Cooper pair is emitted on the same side in the bottom layer. This is the only way for a Cooper pair to enter the condensate. 

The factor two in the denominator of the expression $(\varphi_T-\varphi_B)/2$ can be explained as following. We begin by observing that a $2\pi$-change in the phase difference corresponds to the 
tunneling process of a Cooper pair from one side of an ordinary Josephson junction to the other. The number of Cooper pairs is not conserved and the state of the system is unchanged, \emph{ i.e.} it has the same free energy and current. For our four-terminal junction this process is {\it forbidden} since the motion of a Cooper pair in one layer must be balanced by a counterpropagating pair in the bottom layer. Indeed, if we change $\varphi_T$ to $\varphi_T+2\pi$ we put the junction in a different current state. However, a simultaneous change of $\varphi_B$ by $2\pi$ leaves the junction state unaltered. Thus the fact that the current is a $2\pi$-periodic function of $(\varphi_T-\varphi_B)/2$ and not of {\it e.g.} 
$\varphi_T-\varphi_B$ is a consequence of the fact that electrons are transferred through such a hybrid junction in groups of four.

The peculiar phase dependence we have found has several interesting physical implications. When the top and bottom junctions are polarized with the same phase bias $\varphi_{T} = \varphi_{B}$ (parallel flow) no supercurrent can flow through the EC. In this case the Josephson 
currents experience an {\em exciton blockade}. In the opposite case of counterflow phase bias ($\varphi_{T} = -\varphi_{B}$) the Josephson 
current flowing through the EC is maximal, with a critical value equal to half the critical current of a ballistic one-channel  S-N-S junction. The four-terminal device allows to realize a superdrag effect, {\it i.e.} a drag of dissipationless currents.  When current flows 
in one layer due to a phase bias applied to that layer, a current equal in magnitude but opposite in direction flows in the other layer. This is a consequence of perfect conversion of exciton current into supercurrent.  

At finite temperatures, as long as the EC gap is larger than $\kb T$ the current is essentially unaffected by thermal 
fluctuations. Note that  this occurs even when the thermal length $L_{\text{th}}$ is smaller than  the length $L$ of the junction. This is in striking 
contrast with the case of an ordinary S-N-S junction (or with the case of two decoupled layers), where the critical current is exponentially 
suppressed~\cite{bardeen_johnson_1972} when $L_{\text{th}}\ll L$, due to thermal decoherence affecting a single Andreev-reflection process. 
In the presence of an EC, Andreev reflection processes coherently occurring at the two interfaces transform Cooper pairs into electron-hole 
pairs of the EC, which are protected from thermal decoherence by the excitonic gap.  Thus in the temperature window $\hbar\vf/L \ll \kb T
 \ll |\Gamma|$ the EC counterflow channel is responsible for an \emph{exponential enhancement} of the critical  current.  This effect should be readily observable as an anomalous persistence of the saw-tooth Josephson current as temperature is increased.

\subsection{Realization of the device}

Let us now discuss about possible implementations of the proposed setup. Electron-hole bilayers have been realized in semiconductor ({\it e.g.} GaAs) double quantum wells separated by a thin ({\it e.g.} AlGaAs) barrier. In these systems it is possible to selectively contact one layer by depleting the other one through suitable gating~\cite{Eisenstein_APL_1990}. This technique has been successfully applied in ``Coulomb drag" experiments, which have recently provided indications of EC formation~\cite{croxall2008,seamons2009,sivan1992}. In these systems, however, it may not be easy to contact superconducting electrodes and to measure equilibrium currents, because a large normal gap would arise at the depleted layer between the superconductor and the EC. So far, the observation of exciton condensation under equilibrium conditions has been achieved only in
quantum Hall bilayers at total filling factor $\nu_T=1$. However, quantum Hall systems necessarily have current-carrying gapless channels localized at their edges, which may alter the physics we have discussed above.  

Another possibility is to realize ECs by employing two closely-spaced decoupled graphene layers~\cite{graphene_review_1}, hosting a gas of spatially separated electron-hole pairs.  As compared to semiconductor bilayers, such ECs have been predicted to exhibit extremely high critical temperatures~\cite{min_prb_2008,joglekar_prb_2008,lozovik_jept_2008,mink_prb_2011,kharitonov} and much larger electron and hole densities. Moreover, the small distance between the carrier layers, the weaker dielectric screening, and the linearly-dispersive  bands help to increase both interaction and disorder energy scales.  Furthermore, graphene bands are nearly perfectly particle-hole symmetric, guaranteeing the nesting between the  Fermi surfaces of the conduction and valence bands, and favoring the emergence of a coherent EC state. Josephson currents flowing through graphene contacted to superconducting electrodes 
have already been observed by several groups~\cite{morpurgo_2007}. For these reasons, the use of double-layer graphene seems to be an extremely promising direction. Due to their ultra-high mobilities (even at room temperature), double-layer graphene sheets embedded in a matrix of BN layers~\cite{ponomarenko} are at the moment the most promising graphene-based candidates for the observation of EC in the absence of an external magnetic field. Inter-layer tunneling can be suppressed by interposing a sufficient number of BN layers between the two graphene sheets.

A third realization scheme could be based on 3D Topological Insulators (TIs)~\cite{TIreviews}. These recently discovered materials exhibit Dirac-like conducting surface states separated by an insulating bulk. Recently, it has been put forward that ECs could be realized, at least in principle, by oppositely gating the surfaces of a TI thin film~\cite{seradjeh_2009}, or by inserting a thin insulating layer between the top surface of a TI and the bottom surface of another TI~\cite{wang_2011}. The problem of contacting the surfaces to superconducting electrodes has not been addressed yet. Nevertheless, this type of implementation may become realistic in the near future, in view of the rapid technological advances stimulated by the topological protection offered by these materials.

\subsection{Possible applications}

The unique properties of the conversion of exciton onto Cooper pair supercurrents can be exploited for a number of possible applications.  
In Section~\ref{protected} we discussed how the appearance of an exact double periodicity of the free energy in the 
circuit  shown in Fig.~\ref{circuit}, allows one to realize topologically protected qubits.  It is also possible to imagine a
 device in which two superconducting electrodes contacted to (say) the top layer are enclosed to form a ring-shaped rf SQUID geometry,
so that the phase difference $\varphi_T$ is directly related to the  magnetic flux $\varphi_T=2 \pi \Phi/\Phi_0 +2 \pi n$.
In response to a magnetic field, an induced  Josephson current $I_T$ flows in the top layer  and, according to 
Eq.(\ref{eq:to}), an opposite current $I_B=-I_T$ flows in the bottom layer. Whenever the magnetic field changes  the flux by a 
fluxon $\Phi_0$, the   currents in both layers are reversed.  Current sign switches detected in the bottom layer count the 
fluxons present in the top layer ring. If the magnetic flux is generated by a monotonous analog input signal, the system effectively 
converts it into a sum of current switch pulses, {\it i.e.} to a digital signal.  The system is therefore an analog-to-digital converter. A 
generalization to non-monotonous input signals can easily be achieved by using two double junctions.
Another possible application is photodetection. Indeed the excitonic current is altered if the layers are exposed to an electromagnetic 
or noise source,  resulting in a modification of the Josephson currents in the two junctions.
Finally, we also observe that if electrodes in both layers are enclosed in a ring-shaped geometry, any excitonic supercurrent reversal generates a 
fluxon-antifluxon pair, indicating that this device can realize Josephson fluxon-antifluxon transistors. 

\acknowledgments
We would like to acknowledge a fruitful collaboration with Diego Rainis at the beginning of this activity.
This work has been supported by  FIRB IDEAS project ESQUI, 
EU (programmes NANO-CTM,  IP-SOLID, STREP-QNEMS, STREP-GEOMDISS),  Welch Foundation
Grant No. TBF1473, DOE Division of Materials Sciences and Engineering Grant No. DEFG03-02ER45958, 
SWAN NRI program,  ARO Grant No. W911NF-09-1-0395,  and DARPA Grant No. HR0011-09-1-0009.

%%%%%%%%%%%%%%%%%%%%%%%%%%%%%%%%%%%%%%%%%%%%%%%%%%%%%%%%%%%%%%%%%%%%%%%%%%%%%%%%%%%%%%%%%%%%%%%%%%%%%%%%%%%%%%%%%%%%%%%%%%%%%%%%%%%%

\appendix

\section{Solutions in uniform pairing potentials}\label{app:sol}
In order to calculate the energy density one needs to calculate the transfer matrix which in turn can be  constructed from the eigenfunctions obtained by piecewise solving Hamiltonian \eqref{eq:ham} in the separate regions where the pairing potentials are constant.

\subsection{Exciton Hamiltonian}

Let us start with the excitonic Hamiltonian in the middle region
\begin{equation}
 \mathcal{H}_{\text{\tiny EC},\uparrow} = 
\begin{pmatrix}
 -ip\hbar\vf\partial_x & \Gamma(x) \vs \\
  \Gamma^*(x) & ip\hbar\vf\partial_x
\end{pmatrix} \quad \Gamma(x) = |\Gamma|e^{i(\gamma+qx)}
\end{equation}
Consider first solutions with energy $|\varepsilon -p\hbar\vf q/2|>|\Gamma|$ and wavevector $k$.
Introduce the following very useful notation
\begin{gather}
\label{eq:cos} \cosh\theta \equiv \sqrt{1+\left(\frac{\hbar\vf k}{|\Gamma|}\right)^2}\,,\\
\label{eq:sin} \sinh{\theta} \equiv \frac{\hbar\vf k}{|\Gamma|}\,.
\end{gather}
Then the solutions can be written
\begin{equation}
 \psi^{(r)}_{p}(x) = \begin{pmatrix}
                       \psi_{Tp}^{(r)}(x) \vs \\
		       \psi_{Bp}^{(r)}(x)
                      \end{pmatrix} = 
			\begin{pmatrix}
			 e^{pr\theta/2}e^{i(\gamma+qx)/2} 	\vs\\
			 re^{-pr\theta/2}e^{-i(\gamma+qx)/2}
			\end{pmatrix}e^{ikx}\,,
\end{equation}
where $r$ is a sign defined as
\begin{equation}\label{eq:r}
 r = \mathrm{sign}\left(\eg -p\hbar\vf q/2\right)\,.
\end{equation}
They have dispersion
\begin{equation}
 \varepsilon(k) = p\hbar\vf q/2 \pm \sqrt{|\Gamma|^2+(\hbar\vf k)^2}\,.
\end{equation}

Let us now analyze the case $|\eg -p\hbar\vf q/2|<|\Gamma|$. These solutions can be obtained from the ones found previously upon substituting
\begin{equation}\label{eq:ex_trans}
 k\to i\kappa\quad\text{and}\quad \theta \to i\theta\,,
\end{equation}
and updating the definitions \eqref{eq:cos} and \eqref{eq:sin}
\begin{gather}
 \cos\theta \equiv \sqrt{1-\left(\frac{\hbar\vf \kappa}{|\Gamma|}\right)^2}\,,\\
 \sin{\theta} \equiv \frac{\hbar\vf \kappa}{|\Gamma|}\,.
\end{gather}
We should also include the corresponding solutions for the down-spin wavefunctions, but they can be obtained simply by noting that
\begin{equation}
 \mathcal{H}_{\text{\tiny EC},\downarrow} = -\mathcal{H}_{\text{\tiny EC},\uparrow}^*\,.
\end{equation}

\subsection{Superconducting Hamiltonian}
The Hamiltonian that describes a bulk superconducting contact in the top layer is
\begin{equation}\label{eq:ham_sup}
 \mathcal{H}_S = \begin{pmatrix}
 -ip\hbar\vf\partial_x & \Delta e^{i\varphi} \vs \\
  \Delta e^{-i\varphi} & ip\hbar\vf\partial_x
\end{pmatrix}
\end{equation}
The solutions are almost identical to the ones written previously, so we are just going to fix the notation.
We need only the solutions with positive energy. 
Using 
\begin{gather}
 \cosh\beta \equiv \sqrt{1+\left(\frac{\hbar\vf k_s}{\Delta}\right)^2} \\
 \sinh\beta \equiv \frac{\hbar\vf k_s}{\Delta}
\end{gather}
we have the solutions ($\eg > \Delta$)
\begin{equation}\label{eq:sup_t}
 \psi_{Tp}(x)=
\begin{pmatrix}
  \psi_{T\uparrow p}(x) \vs \\ \psi_{T\downarrow \bar{p}}(x)
 \end{pmatrix} = 
 \begin{pmatrix}
  e^{i\varphi/2}e^{p\beta/2} \vs \\ e^{-i\varphi/2}e^{-p\beta/2}
 \end{pmatrix}e^{ik_s x}\,.
\end{equation}
The dispersion is
\begin{equation}
 \eg(k_s) = \sqrt{\Delta^2+(\hbar\vf k_s)^2}\,.
\end{equation}

In the case $0<\eg<\Delta$ the evanescent wave solutions are again obtained with the substitutions
\begin{equation}\label{eq:sup_trans}
 k_s\to i\kappa_s \text{ and } \beta\to i\beta\,,
\end{equation}
from which it follows
\begin{gather}
 \cos\beta \equiv \sqrt{1-\left(\frac{\hbar\vf \kappa_s}{\Delta}\right)^2}\,, \\
 \sin\beta \equiv \frac{\hbar\vf \kappa_s}{\Delta}\,.
\end{gather}
When considering the bottom layer superconducting contacts one needs to note that the Hamiltonian is given by Eq.~\eqref{eq:ham_sup}
with $p\to -p$.

\section{Diagonalization of the Folded Hamiltonian}
\label{folded-derivation}  
We now give a derivation of equation (\ref{TF-current}).
Start with the field operator written in the form (using the folded Hamiltonian formalism)
\begin{widetext}
\begin{gather}\label{solutions}
 \Psi_{T\sigma +}(x) = \frac{1}{\sqrt{L}}\sum_{r=\pm}\sum_{k}\sqrt{F(rk)}e^{i(k+q/2)x}\hat{c}_{kr} \\
 \Psi_{B\sigma +}(x) = \frac{1}{\sqrt{L}}(-1)^{m_{\gamma}}e^{-i(\varphi_{T,L}-\varphi_{B,L})/2}\sum_{r=\pm}r\sum_{k}\sqrt{F(-rk)}e^{i(k-q/2)x}\hat{c}_{kr}\label{solutions1}
\end{gather}
$m_{\gamma}$ is defined by $(\varphi_{T,L}-\varphi_{B,L})/2-\gamma = m_{\gamma}\pi$ and the sign $r=+$ corresponds to the upper branch (positive energy with respect to $\ef$) of the
dispersion, while $r=-$ to the lower branch. Under the condition $\hbar\vf q \ll |\Gamma|$ one has that the occupancy is purely determined by the sign $r$ of the branch, namely
$\langle\hat{c}_{rk}^\dagger\hat{c}_{rk}\rangle = \delta_{r,-}$ and $\langle\hat{c}_{rk}\hat{c}^\dagger_{rk}\rangle = \delta_{r,+}$.
The expression for the current in layer $\alpha = \pm$ reads 
\begin{equation}
 \hat{I}_{\alpha}(x) = \alpha e \vf\sum_{\sigma = \uparrow,\downarrow}\left(\Psi^\dagger_{\alpha\sigma+}(x)\Psi_{\alpha\sigma+}(x)
-\Psi^\dagger_{\alpha\sigma-}(x)\Psi_{\alpha\sigma-}(x)\right)\,.
\end{equation}
Using the folded Hamiltonian one can rewrite
\begin{equation}
\begin{split}
 \hat{I}_{\alpha}(x) &=  \alpha e \vf\sum_{\sigma = \uparrow,\downarrow}\left(\Psi^\dagger_{\alpha\sigma+}(x)\Psi_{\alpha\sigma+}(x)
-\Psi_{\alpha\sigma+}(-x)\Psi^\dagger_{\alpha\sigma+}(-x)\right)\\ & \equiv \alpha e \vf \sum_{\sigma = \uparrow,\downarrow}
\lim_{y\to x}\left(\Psi^\dagger_{\alpha\sigma+}(x)\Psi_{\alpha\sigma+}(y)
-\Psi_{\alpha\sigma+}(-y)\Psi^\dagger_{\alpha\sigma+}(-x)\right)
\end{split}
\end{equation}
where in the last line we have introduced the correct \emph{point-splitted} definition of the product of two field operators.
The evaluation of the (free) energy and consequently of the supercurrent has to be performed by adopting a point-splitting 
procedure in order to resolve the ill-defined product of the field operators at the same point in space.
This requirement is necessary because of the linearization procedure around the Fermi points.
Introducing now the solution for the fields \eqref{solutions} and \eqref{solutions1} one has
\begin{equation}
\begin{split}
 I_{\alpha}(x) &= \langle \hat{I}_{\alpha}(x)\rangle = \alpha e\vf\sum_{\sigma =  \uparrow,\downarrow}\lim_{y\to x}
\left[\frac{1}{2L}\sum_{r=\pm}\sum_kF(\alpha rk)\left(e^{-i(k+\alpha q/2)(x-y)}\langle\hat{c}_{kr}^{\dagger}\hat{c}^{}_{kr}\rangle
-e^{i(k+\alpha q/2)(x-y)}\langle \hat{c}^{}_{kr}\hat{c}_{kr}^\dagger\rangle\right)\right] 
 \\
 &=\alpha \frac{e\vf}{L}\lim_{y \to x}\sum_{k} \bigg[F(-\alpha k)\left(e^{-i(k+\alpha q/2)(x-y)}f(E^-(k))-e^{i(k+\alpha q/2)(x-y)}f(-E^-(k))\right)
 \\ &+F(\alpha k)\left(e^{-i(k+\alpha q/2)(x-y)}f(E^+(k))-e^{i(k+\alpha q/2)(x-y)}f(-E^+(k))\right) \bigg]
\end{split}
\end{equation}
where $f$ is the Fermi function and $E^r(k) = \hbar\vf q/2+r\sqrt{(\hbar\vf k)^2+|\Gamma|^2}$ the dispersion. Manipulation of this latter expression allows to single out the 
factors related to $T=0$ and finite temperature fluctuations
\begin{equation}
 \begin{split}
  I_{\alpha}(x)= -\alpha\frac{e\vf}{L}\lim_{y\to x}\sum_{k}\left[F(\alpha k)e^{i(k+\alpha q/2)(x-y)} -F(-\alpha k)e^{-i(k+\alpha q/2)(x-y)}\right] \\
  +2\alpha\frac{e\vf}{L}\sum_k\left[F(\alpha k)f(E^+(k))-F(-\alpha k)f(-E^-(k))\right]
 \end{split}
\end{equation}
In the folded geometry the allowed wavevectors have the following form ($J$ is an integer)
\begin{equation}
 k = k^{(0)}+\bar{k}\qquad\qquad k^{(0)} = \frac{\pi n_{odd}}{2L}\qquad\qquad \bar{k} = \frac{\varphi_T+\varphi_B+2\pi J}{4L}\,. 
\end{equation}
Let us study first the ground state contribution
\begin{equation}
 I_{\alpha,GS}(x) = -\alpha\frac{e\vf}{L}\lim_{y\to x}\left[\sum_{k^{(0)}}\left(e^{i\alpha q(x-y)/2}
F(k^{(0)}+\bar{k})e^{i\alpha(k^{(0)}+\alpha\bar{k})(x-y)}
-e^{-i\alpha q(x-y)/2}F(k^{(0)}-\bar{k})e^{i\alpha(k^{(0)}-\alpha\bar{k})(x-y)}\right)\right]
\end{equation}
in the case that $\xi_{ex}^{-1}\gg \Delta k = \Delta_0 = 2\pi/2L= \pi /L$, or in other words $\xi_{ex}\ll L$, the function $F(k)$ appearing in the above equation varies smoothly
with respect to the discrete $k^{(0)}$ spectrum and one can transform the sum into an integral
\begin{equation}
 I_{\alpha,GS}(x) = -\alpha \frac{e\vf}{L}\lim_{y\to x}\left[\int_{-\infty}^{+\infty}dk^{(0)}
\,\left(e^{i\alpha q(x-y)/2}F(k^{(0)}+\bar{k})e^{i\alpha(k^{(0)}+\alpha\bar{k})(x-y)}
-e^{-i\alpha q(x-y)/2}F(k^{(0)}-\bar{k})e^{i\alpha(k^{(0)}-\alpha\bar{k})(x-y)}\right)\right]
\end{equation}
Introducing new integration variables $p = k^{(0)}+\alpha\bar{k}$ in the first term and $p=k^{(0)}-\alpha\bar{k}$ in the second term, 
we notice that the dependence on $\bar{k}$
(\textit{i.e.} on the sum $\varphi_T+\varphi_B$) disappears, whereas the dependence on $q$ (\textit{i.e.} 
on the difference $\varphi_T-\varphi_B$) remains. One thus obtains
\begin{equation}
 \begin{split}
  I_{\alpha,GS}(x) &= -\alpha \frac{e\vf}{\pi}\lim_{y\to x}\left[ e^{i\alpha q(x-y)/2}\left(\int^{+\infty}_{-\infty}dp\,F(p)e^{i\alpha p(x-y)}\right)
-e^{-i\alpha q(x-y)/2}\left(\int^{+\infty}_{-\infty}dp\,F(p)e^{i\alpha p(x-y)}\right)\right]= \\
 &= -2\alpha\frac{e\vf}{\pi} \lim_{y\to x} \left[ \frac{\sin[q(x-y)/2]}{(x-y)} \left(-\int^{+\infty}_{-\infty}dp\,\frac{\partial F(p)}{\partial p}
 \left(e^{i\alpha p(x-y)}\right)\right)\right]
 \end{split}
\end{equation}
We notice that since $F(p)$ is bounded $0\leq F(p) \leq 1$, the integral of its derivative converges, and to lowest order in $x-y$ one can set
\begin{equation}
 \int^{+\infty}_{-\infty}dp\,\frac{\partial F(p)}{\partial p}
 e^{i\alpha p(x-y)} \simeq  \int^{+\infty}_{-\infty}dp\,\frac{\partial F(p)}{\partial p} = F(+\infty)-F(-\infty) = 1
\end{equation}
Thus one obtains
\begin{equation}
 I_{\alpha,GS}(x) = -2\alpha \frac{e\vf}{\pi}\lim_{y \to x}\left[\frac{\sin[q(x-y)/2]}{(x-y)}(-1)\right] =  \alpha \frac{e\vf}{\pi}q ~,
\end{equation}
which (as expected) does not depend on $x$. Recalling that $q$ is given by \eqref{eq:qll} and the integer $n$ must be chosen in order to minimize the 
free energy $\propto q^2$, one has the current in the limit $L\gg \xie$ as given in Eq.~\eqref{eq:to}.

Consider now the finite temperature contribution
\begin{equation}
 \begin{split}
  &I_{\alpha,TF}(x) = -2 \frac{e\vf}{L}\sum_{k^{(0)}}\times \\ &\left[
\frac{F(k^{(0)}-\bar{k})}{1+\exp\left[\beta\left(-\alpha\hbar\vf q/2+\sqrt{|\Gamma|^2+[\hbar\vf(k^{(0)}-\bar{k})]^2}
\right)\right]}-\frac{F(k^{(0)}+\bar{k})}{1+\exp\left[\beta\left(\alpha\hbar\vf q/2+\sqrt{|\Gamma|^2+[\hbar\vf(k^{(0)}+\bar{k})]^2}
\right)\right]}\right]
 \end{split}
\end{equation}
We notice that the above expression depends both on $\bar{k}$ (on the sum $\varphi_T+\varphi_B$) 
and on $q$ (the difference $\varphi_T-\varphi_B$).
Under the conditions $|\Gamma|\gg \hbar\vf q/2$ we observe that the sign of the exponents appearing in the 
Fermi functions is fixed and independent of $k^{(0)}$.
The means that (regardless of the temperature) the integrand varies smoothly with $k^{(0)}$, 
and one can fairly well approximate the sum with an integral,
obtaining
\begin{equation}
 \begin{split}
  &I_{\alpha,TF}(x) = -2 \frac{e\vf}{\pi}\int^{+\infty}_{-\infty}dk^{(0)}\;\times \\ &\left[
\frac{F(k^{(0)}-\bar{k})}{1+\exp\left[\beta\left(-\alpha\hbar\vf q/2+\sqrt{|\Gamma|^2+[\hbar\vf(k^{(0)}-\bar{k})]^2}
\right)\right]}-\frac{F(k^{(0)}+\bar{k})}{1+\exp\left[\beta\left(\alpha\hbar\vf q/2+\sqrt{|\Gamma|^2+[\hbar\vf(k^{(0)}+\bar{k})]^2}
\right)\right]}\right]
 \end{split}
\end{equation}
Each of the two terms separately converges due to the finite temperature. One can than introduce new integration variables 
$p = k^{(0)}-\bar{k}$ in the first term and $p = k^{(0)}+\bar{k}$ in the second term. In doing that, we thus obtain
\begin{equation}
 \begin{split}
  &I_{\alpha,TF}(x) = -2 \frac{e\vf}{\pi}\int^{+\infty}_{-\infty}dp\;\times \\ &\left[
\frac{F(p)}{1+\exp\left[\beta\left(-\alpha\hbar\vf q/2+\sqrt{|\Gamma|^2+(\hbar\vf p)^2}
\right)\right]}-\frac{F(p)}{1+\exp\left[\beta\left(\alpha\hbar\vf q/2+\sqrt{|\Gamma|^2+(\hbar\vf p)^2}
\right)\right]}\right]\,.
 \end{split}
\end{equation}
Furthermore under the condition $\beta|\Gamma| \gg 1$ ($\xie \ll L_{\text{\tiny th}}$) we can approximate
\begin{equation}
 \frac{1}{\sqrt{1+\exp\left[\beta\left(\pm\alpha\hbar\vf q/2+\sqrt{|\Gamma|^2+(\hbar\vf p)^2}
\right)\right]}} \simeq e^{-\beta\left(\pm \alpha \hbar \vf q/2+\sqrt{|\Gamma|^2+(\hbar\vf p)^2}\right)}
\end{equation}
obtaining
\begin{equation}
\begin{split}
 I_{\alpha,TF}(x) &\simeq -2\frac{e\vf}{\pi}\int_{-\infty}^{+\infty}dp\, F(p)
\left[e^{-\beta\left(- \alpha \hbar \vf q/2+\sqrt{|\Gamma|^2+(\hbar\vf p)^2}\right)}-
e^{-\beta\left(\alpha \hbar \vf q/2+\sqrt{|\Gamma|^2+(\hbar\vf p)^2}\right)}\right] = \\
&= -4\frac{e\vf}{\pi}\sinh[\beta\alpha\hbar\vf q/2]\int_{-\infty}^{+\infty}dp F(p)e^{-\beta\sqrt{|\Gamma|^2+(\hbar\vf p)^2}}
\end{split}
\end{equation}
Recalling the definition of the function $F(p)$
\begin{equation}
    F(p) = \frac{1}{2}\frac{\sqrt{|\Gamma|^2+(\hbar\vf p)^2)}+\hbar\vf p}{\sqrt{|\Gamma|^2+(\hbar\vf p)^2)}}
\end{equation}
and noticing that the integrand is even in $p$, one can write
\begin{equation}
 I_{\alpha,TF}(x) = -4\frac{e\vf}{\pi}\sinh[\beta\alpha\hbar\vf q/2]
\int_{0}^{+\infty}dp\frac{\sqrt{|\Gamma|^2+(\hbar\vf p)^2)}+\hbar\vf p}{\sqrt{|\Gamma|^2+(\hbar\vf p)^2)}}e^{-\beta\sqrt{|\Gamma|^2+(\hbar\vf p)^2}}\,.
\end{equation}
The integral can be carried out in terms of the Bessel function $K_1$
\begin{equation}
 I_{\alpha,TF}(x) = -4\alpha\frac{e|\Gamma|}{\pi\hbar}\sinh(\beta\hbar\vf q/2)\left[K_1\left(\frac{L_{\text{\tiny th}}}{\xie}\right)+\frac{\xie
}{L_{\text{\tiny th}}}e^{-\frac{L_{\text{\tiny th}}}{\xie}}\right]\,.
\end{equation}
To first order in $\xie/L_{\text{\tiny th}}$ and using the asymptotic espansion of the Bessel function
\begin{equation}
K_1\left(\frac{L_{\text{\tiny th}}}{\xie}\right)\sim \sqrt{\frac{\pi\xie}{2L_{\text{\tiny th}}}}e^{-\frac{L_{\text{\tiny th}}}{\xie}} 
\end{equation}
 one retrives Eq.~\eqref{eq:finite_temperature_current} (where the ground state contribution has been added). We stress that this result holds
as long as $\xie\ll L,L_{\text{\tiny th}}$.
\end{widetext}

\section{From the scattering matrix to the energy density}\label{app:scatt}
The aim of this section is to offer a simple derivation of Eq.~\eqref{eq:scatt}.
Consider for simplicity a simple scatterer with two ingoing channels ($a_1,a_2$) and two outgoing channels ($b_1,b_2$)
described by a $2\times 2$ scattering matrix $S$
\begin{equation}
 \begin{pmatrix}
  b_1 \\ b_2 
 \end{pmatrix} = 
 \begin{pmatrix}
  r_1 & t_2 \\
  t_1 & r_2
 \end{pmatrix}
 \begin{pmatrix}
  a_1 \\ a_2
 \end{pmatrix}\,.
\end{equation}
The coefficients $r_i$ and $t_i$ are arbitrary function of energy, constrained only by unitarity.
Now suppose that on both sides of the scatterer there are perfectly reflecting mirrors enforcing the following conditions on the wavefunction amplitudes
\begin{gather}
 b_1 = a_1e^{i\phi}\,, \\
 b_2 = a_2e^{i\phi}\,.
\end{gather}
$\phi$ is an arbitrary phase that can be changed at will.
Thus the allowed values of the energy can be found solving the following equation in $\eg$
\begin{equation}\label{eq:eigeneq}
 (e^{i\phi_1(\eg)}-e^{i\phi})(e^{i\phi_2(\eg)}-e^{i\phi}) = 0\,.
\end{equation}
$e^{i\phi_1(\eg)},e^{i\phi_2(\eg)}$ are the eigenvalues of the scattering matrix.
Eq.~\eqref{eq:eigeneq} has in general many solutions since it is satisfied when
\begin{equation}
 \phi_i(\eg) = \phi +2k\pi\,,\quad i= 1,2\,.
\end{equation}
Take $\phi + 2k\pi$ of the form $2\pi n/N$ with $n\in \z$ and with $N$ sufficiently big to have a fine sampling of $\phi_i(\eg)$.
The number of states in the small interval $[\,\eg;\;\eg+\Delta\eg\,]$ is given by
\begin{equation}
 n_i \approx \frac{1}{2\pi}\frac{\partial \phi_i}{\partial \eg} \Delta\eg N\,.
\end{equation}
This expression has a well defined limit for $\Delta\eg \to 0$ and $N\to +\infty$, with their product kept costant, and is proportional to the density of states. In order to determine the  proportionality factor, one can consider the case of a trivial scatterer, i.e. a region of free propagation of a certain length. The right energy density is given by fixing $\Delta\eg N = 1$.
In the end one has
\begin{equation}
 \rho(\eg) = \frac{1}{2\pi}\left(\frac{\partial\phi_1(\eg)}{\partial\eg}+\frac{\partial\phi_2(\eg)}{\partial\eg}\right) \quad,
\end{equation}
an alternative way of writing the more general formula Eq.~\eqref{eq:scatt}.

\section{Scattering matrix determinant}\label{app:det}
We first calculate the block diagonal transfer matrix $T$ of the system 
\begin{equation}
T=\left(
\begin{array}{cc}
T^+ &0\\0&T^-
\end{array}
\right)
\end{equation}
where $T^+$ and $T^-$ are defined by the expression ($d = L/2+M$)
\begin{equation}
\left(\begin{array}{c}
\psi_{T\uparrow p}(d) \\ \psi_{B\uparrow p}(d) \\ \psi_{T\downarrow \bar{p}}(d) \\ \psi_{B\downarrow\bar{p}}(d)
\end{array}\right)=
T^p \left(
\begin{array}{c}
\psi_{T\uparrow p}(-d) \\ \psi_{B\uparrow  p}(-d) \\ \psi_{T\downarrow \bar{p}}(-d) \\ \psi_{B\downarrow\bar{p}}(-d)
\end{array}
\right)\,,
\end{equation}
where $\psi_{\alpha\sigma p}(\pm d)$ are the channels at the two ends ($\pm d$) of the device (with superconducting contacts truncated).
By dividing the system into three regions, one can write that $T^p=T_{\text{\tiny S,R}}^pT_{\text{\ty Ex}}^pT_{\text{\tiny S,L}}^p$, where $T^p_{\text{\tiny S,R}}$ is  related to the right superconducting leads, $T^p_{\text{\tiny S,L}}$ is related to the left superconducting leads, and $T_{\text{\ty Ex}}^p$ is related to the exciton double layer.
 
Once the (unnormalized) eigenfunctions for the three regions are given, their transfer matrices can be written in the form~\cite{titov_2006}
\begin{equation}
 T_{2\times2}(x_2,x_1) = \Lambda(x_2)\begin{pmatrix}
             e^{ik(x_2-x_1)} & 0 \\
	     0 & e^{-ik(x_2-x_1)}
            \end{pmatrix}\Lambda^{-1}(x_1)
\end{equation}
($T_{\text{\tiny  S,(L,R)}}^p$ is block diagonal in the layer space and similarly $T_{\text{\ty Ex}}^p$ is block diagonal in spin space, the above formula is relevant for each $2\times2$ block. $x_2,x_1$ are the two ends connected by the transfer matrix).

$\Lambda$ is the matrix of the wavefunction components reported in Appendix~\ref{app:sol}.
Considering for instance the upper block of $T^p_{\text{\ty Ex}}$ (acting on $\psi_{T\uparrow},\psi_{B\uparrow}$ space) one has
\begin{equation}
\Lambda(x)=
 \begin{pmatrix}
  e^{pr\theta/2}e^{i(\gamma+qx)/2} & e^{-pr\theta/2}e^{i(\gamma+qx)/2} \vs\\
  re^{-pr\theta/2}e^{-i(\gamma+qx)/2} & re^{pr\theta/2}e^{-i(\gamma+qx)/2} 
 \end{pmatrix}\,.
\end{equation}

In particular, for the left (right) superconducting contacts region of length $M$ we find (for the full $4\times4$ matrices)
\begin{widetext}
\begin{equation}\label{eq:trans_l}
T_{S,L(R)}^{p} = \frac{1}{\sinh\beta}\left(
\begin{array}{cccc}
  \sinh \left(\beta +i p M k_s\right) & 0 & -i p e^{i \varphi_{T,L(R)}} \sin \left(M k_s\right) & 0 \vs \\
 0 & \sinh \left(\beta -i p M k_s\right) & 0 & i p e^{i \varphi_{B,L(R)}} \sin \left(M k_s\right) \vs \\
 i  p e^{-i \varphi_{T,L(R)}} \sin \left(M k_s\right) & 0 & \sinh \left(\beta -i p M k_s\right) & 0 \vs \\
 0 & -i p e^{-i \varphi_{B,L(R)}} \sin \left(M k_s\right) & 0 & \sinh \left(\beta +i p M k_s\right) 
\end{array}
\right)\,,
\end{equation}
while for the middle region of the EC double layer of length $L$ we find ($\theta_+,k_+$ is calculated with $p=+$)
\begin{equation}\label{eq:trans_ex}
\begin{split}
 T_{\text{\tiny EC}}^{p} &= \frac{1}{\sinh\theta_p}\times \\ &\left(
\begin{array}{cccc}
 e^{i qL/2} \sinh \left(\theta_p +ipr k_pL\right) & -i p e^{i \gamma}  \sin \left(k_pL\right) & 0 & 0 \vs \\
 i p e^{-i \gamma} \sin \left(k_pL\right) & e^{-i qL/2} \sinh \left(\theta_p -i p rk_pL\right) & 0 & 0 \vs  \\
 0 & 0 & e^{-i qL/2} \sinh \left(\theta_p -i p r k_p L\right) & -i p e^{-i \gamma} \sin \left(k_pL\right) \vs \\
 0 & 0 & i p e^{i \gamma} \sin \left(k_p L\right) & e^{iqL/2} \sinh \left(\theta_p + i pr k_p L\right)
\end{array}
\right)\,,
\end{split}
\end{equation}
\end{widetext}
where $\theta_p$ and $k_p$ are different functions of the energy in each branch
\begin{gather}
 \cosh\theta_p = \left|\frac{\eg -p\hbar\vf q/2}{|\Gamma|}\right|\,, \\
 \sinh\theta_p = \sqrt{\left(\frac{\eg -p\hbar\vf q/2}{|\Gamma|}\right)^2-1}\,,\\
k_p = \frac{|\Gamma|}{\hbar\vf}\sinh\theta_p\,.
\end{gather}

Moreover Eq.~\eqref{eq:trans_l} holds for $\eg > \Delta$, while for $\eg<\Delta$ is sufficient to apply the transformation Eq.~\eqref{eq:sup_trans}. Similarly, Eq.~\eqref{eq:trans_ex} holds for $|\eg| > |\Gamma|$, while for $|\eg|<|\Gamma|$ apply
Eq.~\eqref{eq:ex_trans}.

Now, instead of computing the scattering matrix from the transfer matrix one can prove the following simple formulas for the scattering matrix determinant
\begin{gather}
 \det S^+ = \frac{T^+_{11}T^+_{44}-T^+_{14}T^+_{41}}{T^+_{22}T^+_{33}-T^+_{23}T^+_{32}}\,, \\
 \det S^- = \frac{T^-_{22}T^-_{33}-T^-_{23}T^-_{32}}{T^-_{11}T^-_{44}-T^-_{14}T^-_{41}} \,,
\end{gather}
so that $\det S=\det S^+ \det S^-$.
Here $T^p_{ij}$, with $i,j=1,4$, are the elements of the transfer matrix $T^p$.

Defining the function
\begin{widetext}
 \begin{equation}\label{appi}
\begin{split}
  A_p(\eg,qL,\gamma,\phi_i) = &\bigg(e^{ip\left[qL-\frac{\varphi_T-\varphi_B}{2}\right]}\sinh^2(\theta_p+ik_p L)\sinh^4(\beta+ik_s M) + e^{-ip\left[qL-
\frac{\varphi_T-\varphi_B}{2}\right]}\sinh^2(\theta_p-
ik_p L)\sin^4(k_s M)\\
&+2\sinh^2(\beta+ik_s M)\sin^2(k_s M)\left[\cos[(\varphi_T+\varphi_B)/2]\sinh^2\theta_p+
\cos[2\gamma-(\varphi_L+\varphi_R)/2]\sin^2(k_p L)\right] \bigg) \,,
\end{split} 
\end{equation}
\end{widetext}
where, as before, $k_p\to i\kappa_p$ and $\theta_p\to i\theta_p$ for $|\eg|<|\Gamma|$, $k_s\to i\kappa_s$ and $\beta\to i\beta$ for $\eg<\Delta$.
The scattering matrix determinant is then
\begin{equation}
 \det S = \frac{A_+(\eg,qL,\gamma,\varphi_i)}{A_+^*(\eg,qL,\gamma,\varphi_i)}\cdot\frac{A_-(\eg,qL,\gamma,\varphi_i)}{A_-^*(\eg,qL,\gamma,\varphi_i)}\,.
\end{equation}
Thus the density of states is given by
\begin{equation}\label{eq:intermediate_density}
 \rho(\eg) = \frac{1}{\pi}\frac{\partial}{\partial \eg}\left[\arg A_+(\eg,qL,\gamma,\varphi_i)+\arg A_-(\eg,qL,\gamma,\varphi_i)\right]\,.
\end{equation}

\section{EC free energy}
\label{app:cond_curr}
In the limit $\Delta\to 0$ the expression for $A_p$ in Eq.~(\ref{appi}) reduces to
\begin{equation}
 A_p(\eg,qL) = \sinh^2(\theta_p+irk_pL)
\end{equation}
so that
\begin{equation}\label{eq:density_ex}
\begin{split}
 \rho_{\text{\tiny EC}}(\eg) = \\  = &\sum_{p=\pm}\frac{2}{\pi}\frac{\partial}{\partial\varepsilon}
\arctan\left(\coth\theta_p\tan\left(\tgm\sin\theta_p\right)\right)\\
= &\sum_{p=\pm}\frac{2L}{\pi\hbar\vf}\frac{\cosh^2\theta_p-\frac{\cos(\tgm\sinh\theta_p)\sin(\tgm\sinh\theta_p)}{\tgm\sinh\theta_p}}{\cosh^2\theta_p-\cos^2(\tgm \sinh\theta_p)} \\
\end{split}
\end{equation}
with $\tgm=|\Gamma|L/(\hbar\vf)$.
Note that in the limit of large energy $\eg\gg|\Gamma|$
\begin{equation}\label{rle}
 \lim_{\eg\to+\infty} \rho_{\text{\tiny EC}}(\eg)= \frac{4L}{\pi\hbar\vf} \,,
\end{equation}
while for small energy $\eg\ll|\Gamma|$
\begin{equation}\label{eq:low_eg}
 \lim_{\eg\to 0} \rho_{\text{\tiny EC}}(\eg) \propto \frac{1}{|\Gamma|}\,.
\end{equation}
Because of Eq.~(\ref{rle}), the free energy $\mathcal{F}_{\text{J}}$, defined by Eq.~(\ref{eq:josephson_free_energy}), diverges.
However, it can be regularized by subtracting the constant term associated to the density of states at $q=0$, namely $\tilde{\rho}(\eg)=\rho_{\text{\tiny EC}}(\eg)|_{q=0}$, and redefining the free energy as follows
\begin{equation}\label{eq:free_condensate}
\begin{split}
 \mathcal{F}_{\text{J}}=-\int\limits_0^{+\infty}d\varepsilon\,\varepsilon\left[\rho_{\text{\tiny EC}}(\eg)-\tilde{\rho}(\varepsilon)\right] \,.
\end{split}
\end{equation}

The integral in Eq.~(\ref{eq:free_condensate}) can be evaluated, by introducing a cut-off $\lambda$, as follows 
\begin{equation}
 \begin{split}
 &\mathcal{F}_{\text{J}} =
-\frac{1}{2}\int_0^{\lambda}d\varepsilon\,\varepsilon\tilde{\rho}(\varepsilon-\hbar\vf q/2)-\frac{1}{2}\int_0^\lambda
d\varepsilon\,\varepsilon\tilde{\rho}(\varepsilon+\hbar\vf q/2)\\ &+\int_0^\lambda \varepsilon d\varepsilon\,\tilde{\rho}(\varepsilon)+ O\left(\frac{1}{\lambda^2}\right) = 
\\&=\left(\frac{1}{2}\int\limits_{\lambda-\hbar\vf q/2}^\lambda d\varepsilon\;
  -\frac{1}{2}\int\limits_{\lambda}^{\lambda+\hbar\vf q/2}d\varepsilon\,\;+\int\limits_0^{\hbar\vf q/2} d\varepsilon\,\right)\eg\tilde{\rho}(\varepsilon) + \\ &+\frac{\hbar\vf q}{2}\left(\frac{1}{2}\int\limits_{\lambda -\hbar\vf q/2}^{\lambda+\hbar\vf q/2}d\varepsilon\,-\int\limits_0^{\hbar\vf q/2}d\eg \right) \tilde{\rho}(\varepsilon)+ O\left(\frac{1}{\lambda^2}\right) \,.
\end{split}
\end{equation}
By sending $\lambda$ to infinity, to evaluate the integrals above one needs only the limits in Eqs.~(\ref{rle}) and (\ref{eq:low_eg}) so that
\begin{equation}
\mathcal{F}_{\text{J}} \simeq
\frac{2L}{\pi\hbar \vf}\frac{(\hbar\vf q)^2}{4} + O\left(\frac{1}{\lambda^2}\right) + O\left(\frac{1}{|\Gamma|^2}\right)= \frac{\hbar\vf}{2\pi L}(qL)^2
\end{equation}
This proves Eq.~\eqref{eq:ex_contrib}.

\section{Bound states free energy}\label{app:bound}
Here we derive Eqs.~\eqref{eq:bound_l} and \eqref{eq:bound_r} by matching the solutions of the Bogoliubov-de Gennes equation at the interfaces between the S electrodes and the double layer hosting the EC.
We consider $0<\eg<\Delta$ so that, inside the superconductors, only decaying solutions are present.
Moreover, since $\xi_{\text{\tiny EC}}$ is much smaller than $L$, we consider only decaying solutions inside the EC.
For the left interface, one needs to solve the following set of equations (in the linearized band approximation only the value of the wavefunctions must be matched and not its derivative)
\begin{equation}
 \left\{
\begin{array}{l}
 \psi_{T\uparrow p}(-L/2) = e^{ip\theta}e^{i(\gamma-qL/2)}\psi_{B\uparrow p}(-L/2) \vs \\ 
 \psi_{B\uparrow p}(-L/2) = e^{i\varphi_{B,L}}e^{ip\beta}\psi_{B\downarrow \bar{p}}(-L/2) \vs\\
 \psi_{B\downarrow \bar{p}}(-L/2) = -e^{ip\theta}e^{i(\gamma-qL/2)}\psi_{T\downarrow \bar{p}}(-L/2) \vs\\
 \psi_{T\downarrow \bar{p}}(-L/2) = e^{-i\varphi_{T,L}}e^{ip\beta}\psi_{T\uparrow p}(-L/2)
\end{array}
\right. \,.
\end{equation}
The system has a solution when
\begin{equation}
 \exp i\left(2p\theta+2p\beta+\varphi_{B,L}-\varphi_{T,L}+2\gamma - qL\right) = -1
\end{equation}
{\em i.~e.}
\begin{equation}
 2p\beta +2p\theta = \varphi_L -2\gamma+qL + (2n+1)\pi \,.
\end{equation}
Since $\theta$ is approximately energy-independent when $\varepsilon/|\Gamma|\approx\Delta/|\Gamma|\ll 1$, one obtains
\begin{equation}
 \theta = \arccos\left(-p\frac{q\xie}{2}\right) = \frac{\pi}{2} +p\arcsin\left(\frac{q\xie}{2}\right)
\end{equation}
and therefore
\begin{equation}
 p\beta = \frac{\varphi_L}{2} - \gamma+\frac{qL}{2} - \arcsin\left(\frac{q\xie}{2}\right) + n'\pi \,.
\end{equation}
Finally, recalling that $\cos\beta = \varepsilon/\Delta$,
\begin{equation}\label{eq:b_l}
 \frac{\varepsilon_{L}}{\Delta} = \pm\cos\left[\frac{\varphi_L}{2} - \gamma + \frac{qL}{2} - \arcsin\left(\frac{q\xie}{2}\right)\right]\,,
\end{equation}
where the two solutions correspond to the two possible values for $p$.
For the right interface one has the following set of equations
\begin{equation}
 \left\{
\begin{array}{l}
 \psi_{T\uparrow p}(+L/2) = e^{-ip\theta}e^{i(\gamma+qL/2)}\psi_{B\uparrow p}(+L/2) \vs \\ 
 \psi_{B\uparrow p}(+L/2) = e^{i\varphi_{B,R}}e^{-ip\beta}\psi_{B\downarrow \bar{p}}(+L/2) \vs\\
 \psi_{B\downarrow \bar{p}}(+L/2) = -e^{-ip\theta}e^{i(\gamma+qL/2)}\psi_{T\downarrow \bar{p}}(+L/2) \vs\\
 \psi_{T\downarrow \bar{p}}(+L/2) = e^{-i\varphi_{T,R}}e^{-ip\beta}\psi_{T\uparrow p}(+L/2)
\end{array}
\right.
\end{equation}
and to the related quantization condition
\begin{equation}
 \exp i\left(-2p\beta-2p\theta +2\gamma+qL+\varphi_{B,R}-\varphi_{T,R}\right) = -1 \,.
\end{equation}
Finally, one gets
\begin{equation}\label{eq:b_r}
 \frac{\varepsilon_R}{\Delta} = \pm\cos\left[\frac{\varphi_R}{2}-\gamma - \frac{qL}{2} +\arcsin\left(\frac{q\xie}{2}\right)\right] \,.
\end{equation}\\

Eqs.~(\ref{eq:bound_l}) and (\ref{eq:bound_r}) are obtained by observing that the arcsines in the cosine arguments of Eqs.~\eqref{eq:b_l} and \eqref{eq:b_r} are small and amount to a simple renormalization of the junction length due to the finite penetration (of order $\xie$) of the electron wavefunctions in the EC and will be omitted in the following. 

%\vspace*{0.5cm}
%\begin{widetext}
An alternative derivation of the two bound states $\eg_{L}$ and $\eg_{R}$ can be performed by using Eq.~(\ref{eq:intermediate_density}) to calculate the density of states  $\rho_{\text{\tiny BS}}$.
In the limit $|\Gamma|\to \infty$, Eq.~(\ref{appi}) reduces to
\begin{widetext}
\begin{equation}\label{eq:arg_sup}
\begin{split}
A_p(\eg,qL,\gamma,\varphi_i) &= 
 \bigg(e^{ip[qL-(\varphi_T-\varphi_B)/2]}\sin^4(\beta+i k_s M) 
+e^{-ip[qL-(\varphi_T-\varphi_B)/2]}\sinh^4(k_s M)\\ 
&+2\sin^2(\beta+i k_s M)\sinh^2 k_s M\cos[2\gamma-(\varphi_L+\varphi_R)/2]\bigg)\,.
\end{split}
\end{equation}
Using that fact that 
\begin{equation}
\sin(\beta+ik_s M) = e^{i\alpha}\sqrt{\sin^2\beta+\sinh^2 k_s M}
\end{equation}
and defining
\begin{equation}
\alpha = \arctan (\cot\beta\tanh k_s M) = \arctan (\tan(\pi/2-\beta)\tanh k_s M) \,,
\end{equation}
in the limit $M\to +\infty$ one finds that
\begin{equation}\label{eq:up}
 \begin{split}
 \varphi_+(\alpha) \equiv \arg A_+(\eg,qL,\gamma,\varphi_i) = \arctan\left(\frac{2\delta\sin[qL-(\varphi_T-\varphi_B)/2+2\alpha]}
{(2+\delta^2)\cos[qL-(\varphi_T-\varphi_B)/2+2\alpha]+
2\cos[2\gamma-(\varphi_L+\varphi_R)/2]}\right)
 \end{split}
\end{equation}
\end{widetext}
where
\begin{equation}
\delta = \frac{\sin^2\beta}{\sinh^2k_sM}
\end{equation}
is a small quantity and can be taken energy independent in the limit $M\to \infty$.
In Eq.~(\ref{eq:up}) we have retained term up to first order in $\delta$ in the numerator and up to second order in the denominator.
The only energy dependent quantity left in Eq.~\eqref{eq:up} is $\alpha(\eg)$. Instead of calculating the density of states as a function of energy is much more convenient to use $\alpha$ as the independent variable since what matters is the invariant measure $(\partial\rho(\alpha)/ \partial\alpha) d\alpha = (\partial \rho(\eg)/ \partial\eg)d\eg$.
Computing the derivative to the same approximation in $\delta$ we obtain
\begin{equation}\label{eq:delta_func}
\frac{\partial \varphi_+(\alpha)}{\partial\alpha} \approx \frac{2\delta\left(1+\cos\Phi_1\cos\Phi_2\right)}
{\left(\cos\Phi_1+\cos\Phi_2\right)^2+\delta^2\left(1+\cos\Phi_2\cos\Phi_1\right)}
\end{equation}
where
\begin{equation}
\Phi_1=qL-\frac{\varphi_T-\varphi_B}{2}+2\alpha
\end{equation}
and
\begin{equation}
\Phi_2= 2\gamma-\frac{\varphi_L+\varphi_R}{2} \,.
\end{equation}
an identical calculation for $p=-$ gives the same result with $\alpha\to-\alpha$.

This function is strongly peaked when the zero order term in the denominator vanishes.
In the limit $M\to+\infty$ we have $\alpha\to \pi/2 -\beta$, so
\begin{equation}
 -\cos[qL-(\varphi_T-\varphi_B)/2+2p\beta]+\cos[2\gamma-(\varphi_L+\varphi_R)/2] = 0
\end{equation}
in other words
\begin{equation}
  2p\beta =\left\{\begin{array}{c}
                   2\gamma-qL -\varphi_L \\
                   2\gamma+qL -\varphi_R
                  \end{array}\right.
\end{equation}
so we have found the energies of two bound states
\begin{gather}
  \varepsilon_L = \pm\Delta\cos\left(\gamma-qL/2-\varphi_L/2 \right)\,, \\
 \varepsilon_R = \pm\Delta\cos\left(\gamma+qL/2-\varphi_R/2\right)\,.
\end{gather}
We can speak about bound states since, by expanding  around $\eg_R$ or $\eg_L$, one can see that in the limit $M\to +\infty$ 
Eq.~\eqref{eq:delta_func} tends to a delta function with weight $\pi$.

We have also verified that the contribution for $\eg>\Delta$ of 
Eq.~\eqref{eq:arg_sup} washes out for $M\to +\infty$. The only relevant features of the low energy spectrum are the two bound state peaks.


\begin{thebibliography}{77}
\bibitem{nota-on-L} 
	While in the folded Hamiltonian method (Sec.~\ref{longjunctionsec}) it is particularly suitable to set the coordinates 
	of the S-EC interfaces at $x=0$ and to $x=L$, 
	in the truncation method it is easier to exploit the symmetry of the setup by setting the interfaces at $x=\pm L/2$.
\end{thebibliography}
\end{document}